\documentclass[aps,prl,twocolumn,preprintnumbers,amsmath,amssymb,superscriptaddress,10pt]{revtex4-2}

\usepackage[utf8]{inputenc}
\usepackage{graphicx}
\usepackage{bm}

\usepackage{color}
\definecolor{darkblue}{rgb}{0,0,0.6}

\usepackage{ifpdf}
\ifpdf
\usepackage{epstopdf}
\usepackage[pdftex,unicode,pdfstartview={FitH},pdfborder={0 0 0}]{hyperref}
\usepackage{hypcap}
\else
\usepackage[hypertex]{hyperref}
\fi
\hypersetup{
    bookmarksnumbered = true,
    colorlinks = true, linkcolor = darkblue,
    citecolor = darkblue, filecolor = darkblue,
    menucolor = darkblue, urlcolor = darkblue
}

\usepackage{dcolumn}
\newcolumntype{R}{>{$\displaystyle}r<{$}}
\newcolumntype{C}{>{$\displaystyle}c<{$}}
\newcolumntype{L}{>{$\displaystyle}l<{$}}

\begin{document}
    
\title{Field-induced hybridization of moir\'e excitons in MoSe$_2$/WS$_2$ heterobilayers}

\author{Borislav Polovnikov}
\email{borislav.polovnikov@physik.lmu.de}
\affiliation{Fakult\"at f\"ur Physik, Munich Quantum Center, and Center for NanoScience (CeNS), Ludwig-Maximilians-Universit\"at M\"unchen, Geschwister-Scholl-Platz 1, 80539 M\"unchen, Germany}
\affiliation{Max-Planck-Institut f\"ur Quantenoptik, Hans-Kopfermann-Straße 1, 85748 Garching bei M\"unchen, Germany}
\author{Johannes Scherzer}
\affiliation{Fakult\"at f\"ur Physik, Munich Quantum Center, and Center for NanoScience (CeNS), Ludwig-Maximilians-Universit\"at M\"unchen, Geschwister-Scholl-Platz 1, 80539 M\"unchen, Germany}
\author{Subhradeep Misra}
\affiliation{Fakult\"at f\"ur Physik, Munich Quantum Center, and Center for NanoScience (CeNS), Ludwig-Maximilians-Universit\"at M\"unchen, Geschwister-Scholl-Platz 1, 80539 M\"unchen, Germany}
\author{Xin Huang}
\affiliation{Fakult\"at f\"ur Physik, Munich Quantum Center, and Center for NanoScience (CeNS), Ludwig-Maximilians-Universit\"at M\"unchen, Geschwister-Scholl-Platz 1, 80539 M\"unchen, Germany}
\affiliation{Beijing National Laboratory for Condensed Matter Physics, Institute of Physics, Chinese Academy of Sciences, Beijing 100190, People’s Republic of China}
\affiliation{School of Physical Sciences, CAS Key Laboratory of Vacuum Physics, University of Chinese Academy of Sciences, Beijing 100190, People’s Republic of China}
\author{Christian Mohl}
\affiliation{Fakult\"at f\"ur Physik, Munich Quantum Center, and Center for NanoScience (CeNS), Ludwig-Maximilians-Universit\"at M\"unchen, Geschwister-Scholl-Platz 1, 80539 M\"unchen, Germany}
\author{Zhijie Li}
\affiliation{Fakult\"at f\"ur Physik, Munich Quantum Center, and Center for NanoScience (CeNS), Ludwig-Maximilians-Universit\"at M\"unchen, Geschwister-Scholl-Platz 1, 80539 M\"unchen, Germany}
\author{Jonas G{\"o}ser}
\affiliation{Fakult\"at f\"ur Physik, Munich Quantum Center, and Center for NanoScience (CeNS), Ludwig-Maximilians-Universit\"at M\"unchen, Geschwister-Scholl-Platz 1, 80539 M\"unchen, Germany}
\author{Jonathan F\"orste}
\affiliation{Fakult\"at f\"ur Physik, Munich Quantum Center, and Center for NanoScience (CeNS), Ludwig-Maximilians-Universit\"at M\"unchen, Geschwister-Scholl-Platz 1, 80539 M\"unchen, Germany}
\author{Ismail Bilgin}
\affiliation{Fakult\"at f\"ur Physik, Munich Quantum Center, and Center for NanoScience (CeNS), Ludwig-Maximilians-Universit\"at M\"unchen, Geschwister-Scholl-Platz 1, 80539 M\"unchen, Germany}
\author{Kenji Watanabe}
\affiliation{Research Center for Functional Materials, National Institute for Materials Science, 1-1 Namiki, Tsukuba 305-0044, Japan}
\author{Takashi Taniguchi}
\affiliation{International Center for Materials Nanoarchitectonics, National Institute for Materials Science, 1-1 Namiki, Tsukuba 305-0044, Japan}
\author{Alexander H{\"o}gele}
\email{alexander.hoegele@lmu.de}
\affiliation{Fakult\"at f\"ur Physik, Munich Quantum Center, and Center for NanoScience (CeNS), Ludwig-Maximilians-Universit\"at M\"unchen, Geschwister-Scholl-Platz 1, 80539 M\"unchen, Germany}
\affiliation{Munich Center for Quantum Science and Technology (MCQST), Schellingstra\ss{}e 4, 80799 M\"unchen, Germany}
\author{Anvar~S.~Baimuratov}
\email{anvar.baimuratov@lmu.de}
\affiliation{Fakult\"at f\"ur Physik, Munich Quantum Center, and Center for NanoScience (CeNS), Ludwig-Maximilians-Universit\"at M\"unchen, Geschwister-Scholl-Platz 1, 80539 M\"unchen, Germany}

\date{\today}

\begin{abstract}
We study experimentally and theoretically the hybridization among intralayer and interlayer moir\'e excitons in a MoSe$_2$/WS$_2$ heterostructure with antiparallel alignment. Using a dual-gate device and cryogenic white light reflectance and narrow-band laser modulation spectroscopy, we subject the moir\'e excitons in the MoSe$_2$/WS$_2$ heterostack to a perpendicular electric field, monitor the field-induced dispersion and hybridization of intralayer and interlayer moir\'e exciton states, and induce a cross-over from type~I to type~II band alignment. Moreover, we employ perpendicular magnetic fields to map out the dependence of the corresponding exciton Land\'e $g$-factors on the electric field. Finally, we develop an effective theoretical model combining resonant and non-resonant contributions to moir\'e potentials to explain the observed phenomenology, and highlight the relevance of interlayer coupling for structures with close energetic band alignment as in MoSe$_2$/WS$_2$.
\end{abstract}

\maketitle
\copyright 2024 American Physical Society\\
Two dimensional transition metal dichalcogenides (TMDs)~\cite{Colloquium_Wang2018,FaugerasPotemski2017,manzeli_2d_2017} and vertically stacked van der Waals heterostructures~\cite{Geim2013} have emerged as increasingly significant materials for condensed matter research. Stacking two atomically thin layers with small lattice mismatch or rotational misalignment generally results in a new, long-range moir\'e superlattice that introduces a periodic potential for charge-carriers and provides a scaffold for ordered electronic states~\cite{superconductivity_flat_bands_2020,Bistritzer2011,mak_semiconductor_2022}. The periodicity of the moir\'e potential depends sensitively on the rotation angle between the layers~\cite{Hermann2012}, providing means to engineer the physical properties of van der Waals structures and to favor the formation of correlated many-body states~\cite{Dean2018_Twistable_Graphene}.

In their monolayer limit TMDs host tightly bound excitons with large binding energy and strong light-matter interaction~\cite{Colloquium_Wang2018}. In twisted or lattice-incommensurate TMD heterobilayers, the presence of a moir\'e potential leads to the emergence of distinct moir\'e excitons that can be used to study electronic states optically~\citep{Wilson2021,huang_excitons_2022}, including signatures of Hubbard model physics and correlation-induced magnetism~\cite{Tang2020,Gerardot_exciton-polarons_2022}, Wigner crystallization of charge carriers~\cite{Regan2020,Zhou2021} or correlated Mott-insulator states~\cite{Regan2020,Shimazaki2020,Wang2020a,Huang2021,Ghiotto2021,Xu2020,LiMak2021mit}. 

The MoSe$_2$/WS$_2$ heterostructure, which has previously been shown to host correlated electronic states~\cite{Mak2021_capacitance_charge_order, Tang2022}, deserves particular analysis. Unlike in WS$_2$/WSe$_2$ or MoSe$_2$/WSe$_2$, which have conduction (CB) and valence band (VB) offsets of a few hundreds of meV, first-principle calculations have shown that the CBs of MoSe$_2$ and WS$_2$ are much closer~\cite{Zhang2016}. However, their precise energetic ordering, and hence the nature of the lowest energy moir\'e excitons in MoSe$_2$/WS$_2$, remain matter of continuing debate, with reported CB offset values between $-20$~meV~\cite{Zhang_Deng_Twist2020} and $+100$~meV~\cite{Tang2022}. Moreover, early studies mostly focused on a parallel (R-type) stacking configuration with opposite spins in the lowest conduction subbands. In this work, by using complementary experimental and theoretical techniques, we present a comprehensive study of antiparallel (H-type)  MoSe$_2$/WS$_2$ that unveils an intricate interplay of both intra- and interlayer moir\'e excitons and allows to unambiguously determine the nature of the observed optical resonances.

%
\begin{figure}[t!]  
\includegraphics[width=\columnwidth]{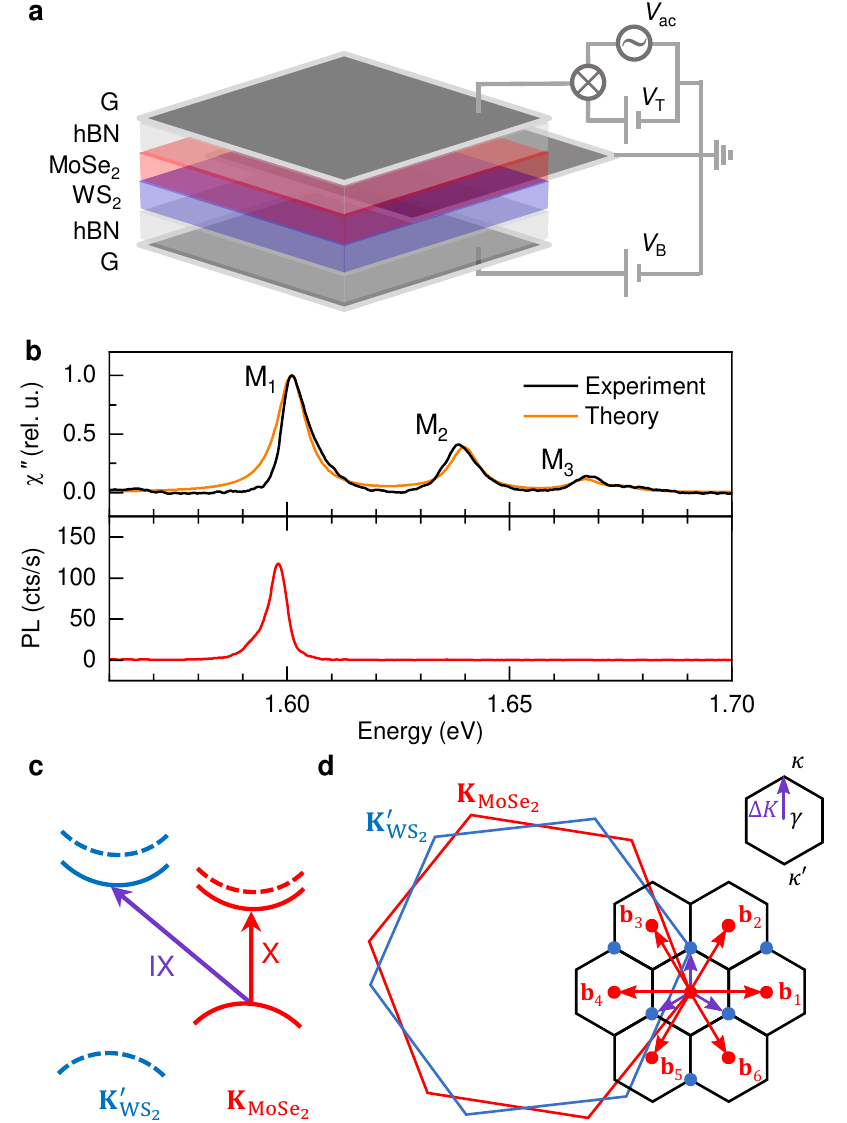}
\caption{\textbf{a}. Schematics of the MoSe$_2$/WS$_2$ heterostructure in a field-effect device, with top and bottom gate voltages $V_\text{T}$ and $V_\text{B}$ and a small ac-voltage $V_\text{ac}$ for narrow-band laser modulation spectroscopy. \textbf{b}. Top panel: Moir\'e exciton peaks $M_1$, $M_2$ and $M_3$ in experimental $\chi''$ as obtained from white light reflection spectra (black) shown together with the theoretical fit (orange). Bottom panel: The PL spectrum is dominated by the lowest-energy state $M_1$ due to fast population relaxation. \textbf{c}. Band structure of antiparallel MoSe$_2$/WS$_2$ with spin-polarized subbands indicated by solid and dashed lines, as well as intra- and interlayer excitons. \textbf{d}. Mini Brillouin zones with moir\'e reciprocal vectors $\mathbf{b}_j$ (red arrows) and interlayer coupling (violet arrows). The red and blue points represent intra- and interlayer states considered in the model.}
\label{fig1}
\end{figure}
%

To control the resonance energy of interlayer excitons we fabricated a dual-gate field-effect device with the schematics shown in Fig.~\ref{fig1}a, consisting of a MoSe$_2$/WS$_2$ heterobilayer encapsulated by symmetric hexagonal boron nitride (hBN) spacers of 55~nm each with few-layer graphite electrodes (G) as the top and bottom gates. Applying opposite gate voltages $V_\text{B}$ and $V_\text{T} = -V_\text{B}$ subjects the device to a perpendicular electric field proportional to $\Delta V_\text{TB} = V_\text{B} - V_\text{T}$. By using crystals synthesized with chemical vapor deposition~\cite{bilgin2015chemical}, we controlled the orientation of the monolayers by optically aligning the edges of the flakes and fabricated a sample with a 179$\pm$1° angle~\cite{SuppMat}. In this case, the $K$-valley of MoSe$_2$ is aligned with the $K'$-valley of WS$_2$. In Fig.~\ref{fig1}c we show the spin configuration of the lowest-energy CBs and the highest-energy VBs for the respective valleys of the two layers.

We performed cryogenic white-light differential reflectance (DR) measurements and computed the imaginary part of the dielectric susceptibility $\chi''$ proportional to absorption~\cite{SuppMat}. In Fig.~\ref{fig1}b, we show the spectrum at charge neutrality with three moir\'e exciton peaks $M_1$, $M_2$ and $M_3$. The peak $M_1$ at 1.60~eV is approximately 30 meV below the monolayer MoSe$_2$ A-exciton at 1.63 eV and has the largest oscillator strength. The peaks $M_2$ and $M_3$ have decreasing oscillator strengths and lie 38 and 65 meV above $M_1$, respectively. $M_1$ is the exciton ground state of the system with bright photoluminescence, whereas the higher states are quenched by rapid population relaxation. 

The peaks $M_1$ and $M_2$ are present in both H- and R-type stackings and have been reported in previous works on MoSe$_2$/WS$_2$ \cite{AlexeevTartakovsky2019,Zhang_Deng_Twist2020,Tang2021,Tang2022}. In early studies~\cite{AlexeevTartakovsky2019,Zhang_Deng_Twist2020,RuizFalko2019}, they have been interpreted as hybridized intra-interlayer excitons and successfully modeled via interlayer tunneling~\cite{Yu_IXcoupling_2017,RuizFalko2019}. However, recent experiments on the exciton dispersion in perpendicular electric fields observed vanishing out-of-plane dipoles for both $M_1$ and $M_2$, suggesting that they are of pure intralayer character~\cite{Tang2021,Tang2022}. While this allows to describe the lowest-energy excitons in terms of a continuum model similarly to WS$_2$/WSe$_2$~\cite{MacdonaldTopo2017,MacdonaldIX2018,MacdonaldHubbard2018,TranMacdonald2019, FengWang2019}, the role of interlayer hybridization remains an open question that we address in this work. Therefore, to describe the complexity of both intra- and interlayer excitons in MoSe$_2$/WS$_2$ simultaneously, we first proceed by developing an effective continuum model and subsequently demonstrate its validity by comparison with experimental observations.

We start by developing a phenomenological model combining an intralayer moir\'e potential with near-resonant interlayer hybridization. The large VB offset between MoSe$_2$ and WS$_2$~\cite{Zhang2016} simplifies the analysis of the lowest-energy spin-bright excitons, and it is sufficient to consider only intralayer excitons (X) of MoSe$_2$ and interlayer excitons (IX) consisting of a hole in MoSe$_2$ and an electron in WS$_2$ (cp. Fig.~\ref{fig1}c). The twist angle $\theta \approx 179^\circ$ and the lattice constant mismatch between the two layers result in a small valley mismatch which we denote by $\Delta\mathbf{K} = \mathbf{K}_\text{\footnotesize{WS$_2$}}' - \mathbf{K}_\text{\footnotesize{MoSe$_2$}}$. The long-range periodicity of the moir\'e lattice is reflected by the formation of a mini Brillouin zone (mBZ)~\cite{Hermann2012} with the moir\'e reciprocal lattice vectors $\mathbf{g}(n,m) = n\mathbf{b}_1 + m\mathbf{b}_2$ shown in Fig.~\ref{fig1}d. Here, $n$ and $m$ are integers, $\mathbf{b}_j = (C_6^{j-3}-C_6^{j+1})\Delta \mathbf{K}$  are the first-shell reciprocal lattice vectors, $j = 1,2,...,6$, and $C_\nu^\mu$ represents rotation by $2\pi \mu/\nu$ (Fig.~\ref{fig1}d). We define the angle-dependent mBZ as in Fig.~\ref{fig1}d with the center $\gamma$ matching the $K$-valley of MoSe$_2$ ($\mathbf{K}_\text{\footnotesize{MoSe$_2$}}$), and the point $\kappa$ at the $K'$-valley of WS$_2$ ($\mathbf{K}_\text{\footnotesize{WS$_2$}}'$).

To define the general moir\'e Hamiltonian, we introduce two moir\'e potentials~\cite{MacdonaldTopo2017,MacdonaldIX2018,MacdonaldHubbard2018} for X and IX excitons:
\begin{equation}
\begin{split}
V(\mathbf{r})=&\sum_{j=1}^6 V_j \exp{(i\mathbf{b}_j\mathbf{r})} \, , \\
W(\mathbf{r})=&\sum_{j=1}^6 W_j \exp{(i\mathbf{b}_j\mathbf{r})} \, . \nonumber
\end{split}
\end{equation}
These are the lowest-order harmonic expansions of the moir\'e potential which, due to the 120° rotational symmetry, present the usual symmetry relations $V_1 = V_3 = V_5 \equiv V$ and $V_2 = V_4 = V_6 \equiv V^*$ (and analogously for $W$). Restricting ourselves to the low-energy physics of moir\'e excitons, we assume parabolic dispersions, $E(\mathbf{k}) = E_\text{X} + \hbar^2 |\mathbf{k}|^2 / (2M_\text{X})$ and $\mathcal{E}(\mathbf{k'}) = E_\text{IX} + \hbar^2 |\mathbf{k'}|^2 / (2M_\text{IX})$. Here $\mathbf{k}$ and $\mathbf{k}'$ are the center-of-mass wave vectors of X and IX excitons measured from $\gamma$ and $\kappa$, respectively, $M_\text{X}$ and $M_\text{IX}$ are their effective masses, and $E_\text{X}$ and $E_\text{IX}$ are bandgaps averaged over the moir\'e supercell, resulting in:
\begin{equation}
\begin{split}
    \langle \mathbf{k} + \mathbf{g'}| H_\text{X}| \mathbf{k + g} \rangle_\text{X} =&  \delta_{\mathbf{g,g'}} E(\mathbf{k}) + \sum_{j=1}^6 V_j \delta_{\mathbf{g-g',b}_j} \, ,
    \\
    \langle \mathbf{k'} + \mathbf{g'}| H_\text{IX} | \mathbf{k' + g} \rangle_\text{IX} =& \delta_{\mathbf{g,g'}} \mathcal{E}(\mathbf{k'}) + \sum_{j=1}^6 W_j \delta_{\mathbf{g-g',b}_j} . \nonumber
\end{split}
\end{equation}
Notably, $V$ and $W$ do not include resonant interaction terms between X and IX excitons yet, which we introduce explicitly~\cite{RuizFalko2019} by defining the Hamiltonian as:
 \begin{equation}
    H = \\
    \begin{pmatrix}
      H_\text{X} & T \\
      T^*& H_\text{IX}
    \end{pmatrix}, \nonumber
    \label{Ham}
\end{equation}
where the tunneling is described by interlayer hopping elements
\begin{equation}
    \langle \text{IX}, \mathbf{k' + g'}|
    T|
    \text{X}, \mathbf{k + g} \rangle =
    \sum_{\eta=0}^2 t \delta_{\mathbf{k+g-k'-g'}, C^\eta_3 \Delta \mathbf{K}}, \nonumber
\end{equation}
with the hopping parameter $t$. We follow the approximation introduced in Refs.~\cite{Yu_IXcoupling_2017,RuizFalko2019} and assume the same term $t$ in all mBZs, mediating the hybridization of MoSe$_2$ excitons X with the closest three IX states in the reciprocal space denoted by the violet arrows in Fig.~\ref{fig1}d.
%
\begin{figure}[t]  
\includegraphics[width=\columnwidth]{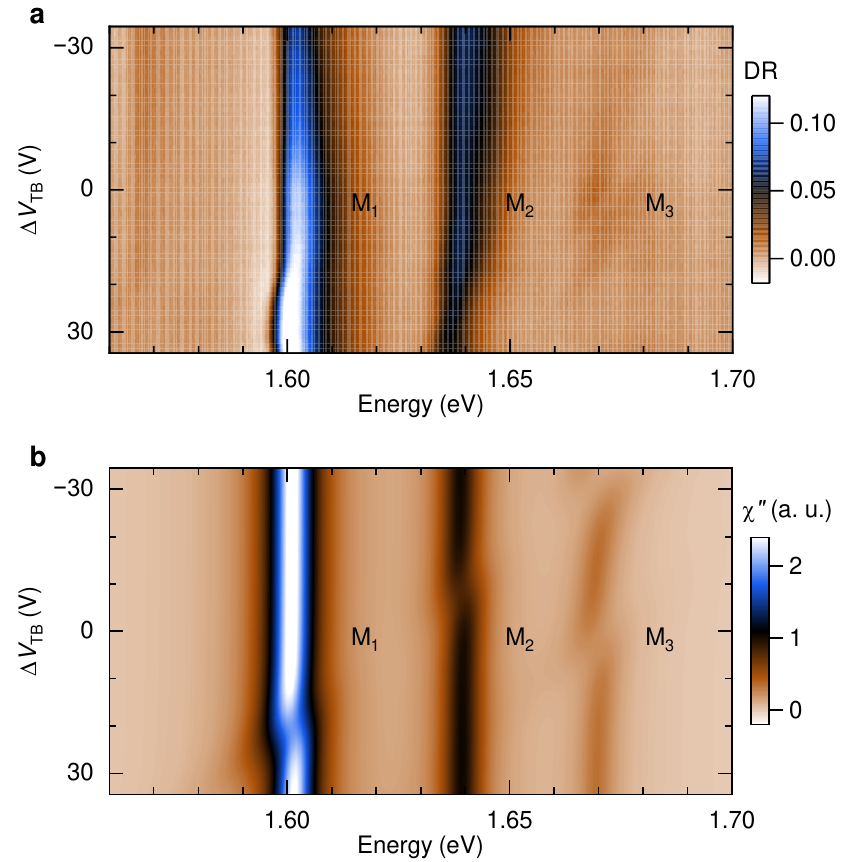}
\caption{\textbf{a}. Experimental DR signal of MoSe$_2$/WS$_2$ as a function of the out-of-plane electric field. \textbf{b}. Corresponding simulation of $\chi'' (\omega)$ capturing the strong intralayer character of the three lowest-energy moir\'e exciton peaks as well as their relative strengths. Inhomogeneous broadening was included by smoothing  $\chi'' (\omega)$ over a $5$~meV broad Gaussian kernel in the fit parameter $E_\text{IX}$.}
\label{fig2}
\end{figure}
%

Finally, we restrict the number of vectors $\mathbf{g}$ considered in the Hamiltonian and keep only the lowest seven X-bands and six IX-bands arising from band folding of mBZs as specified by the red and blue dots in Fig.~\ref{fig1}d, respectively. Furthermore, we assume that the non-resonant part of the interlayer potential is zero, $W_{1...6} = 0$, resulting in the 13-band Hamiltonian:

\begin{equation}\label{HamMatrix}
    H(\mathbf{k}) =
    \left( 
    \begin{array}{lllllllllllll}
    E_0  & V     & V^* & V   & V^* & V   & V^* & t    & t    & t    & 0    & 0    & 0 \\
    V^* & E_1  & V   & 0   & 0   & 0   & V   & 0    & 0    & t    & 0    & t    & 0 \\
    V     & V^* & E_2 & V^* & 0   & 0   & 0   & t    & 0    & 0    & 0    & t    & 0 \\
    V^* & 0     & V   & E_3 & V   & 0   & 0   & t    & 0    & 0    & 0    & 0    & t \\
    V     & 0     & 0   & V^* & E_4 & V^* & 0   & 0    & t    & 0    & 0    & 0    & t \\
    V^* & 0     & 0   & 0   & V   & E_5 & V   & 0    & t    & 0    & t    & 0    & 0 \\
    V     & V^* & 0   & 0   & 0   & V^* & E_6 & 0    & 0    & t    & t    & 0    & 0 \\ 
    t^*  & 0  & t^* & t^* & 0   & 0   & 0   & \mathcal{E}_0 & 0  & 0 & 0 & 0 & 0 \\
    t^*  & 0   & 0 & 0 & t^* & t^* & 0 & 0  & \mathcal{E}_1 & 0    & 0    & 0    & 0 \\
    t^*  & t^* & 0   & 0   & 0  & 0   & t^* & 0  & 0  & \mathcal{E}_2 & 0    & 0    & 0 \\
    0    & 0    & 0  & 0  & 0 & t^* & t^* & 0  & 0  & 0  & \mathcal{E}_3 & 0    & 0 \\
    0   & t^* & t^* & 0  & 0   & 0  & 0  & 0   & 0   & 0   & 0   & \mathcal{E}_4 & 0 \\
    0   & 0   & 0 & t^* & t^* & 0  & 0  & 0   & 0   & 0   & 0   & 0   & \mathcal{E}_5
    \end{array}
	\right)
	.
\end{equation}

The first seven diagonal terms correspond to the X bands marked by the red dots in Fig.~\ref{fig1}d, with
\begin{equation}
    E_0(\mathbf{k}) = E(\mathbf{k})  , \quad E_j(\mathbf{k}) = E (\mathbf{k-b}_j), \nonumber
\end{equation}
for $j=1,2,...,6$; the last six diagonal terms relate to IXs marked by the blue dots in Fig.~\ref{fig1}d:
\begin{equation}
    \mathcal{E}_\eta(\mathbf{k}) = \mathcal{E}(\mathbf{k} - C^\eta_3 \Delta \mathbf{K}) , \quad  \mathcal{E}_\zeta(\mathbf{k}) = \mathcal{E}(\mathbf{k} + 2 C^{\zeta-3}_3 \Delta \mathbf{K}),
\label{IX}
\end{equation}
where $\eta=0,1,2$ and $\zeta=3,4,5$. This Hamiltonian combines the non-resonant intralayer continuum model described by the potential $V$~\cite{MacdonaldTopo2017,MacdonaldIX2018,MacdonaldHubbard2018} with the interlayer hybridization described by the hopping parameter $t$~\cite{Yu_IXcoupling_2017,RuizFalko2019,SuppMat}, and it can be used to compute the band dispersion within the full mBZ. Here, since we are mainly interested in the optical response of the system, we consider the center of the mBZ at the $\gamma$-point. After diagonalization of the Hamiltonian in Eq.~(\ref{HamMatrix}), we obtain $\chi''$ by projecting the eigenstates onto the fundamental A-exciton state as:
\begin{equation}
  \chi'' (\omega) \approx
  \chi''_0
  \sum_{m=1}^{13}
  \left|
    \langle m|A\rangle
  \right|^2 
  \frac{\gamma_0^2}{\hbar^2(\omega-\omega_m)^2 + \gamma_0^2},
  \label{optcond}
\end{equation}
where $|A\rangle$ is the MoSe$_2$ intralayer exciton state corresponding to the first row and column in Eq.~(\ref{HamMatrix}), $\chi''_0$ is its dielectric susceptibility~\cite{MacdonaldTopo2017}, $|m\rangle$ and $\hbar\omega_m$ are the eigenstates and eigenvalues of the $m$-th exciton band obtained from the diagonalized Hamiltonian in Eq.~\eqref{HamMatrix}, and $\gamma_0$ is a peak broadening parameter. Notably, this implies that the oscillator strength of the $|A\rangle$ exciton is redistributed among the set of the moir\'e excitons $|m\rangle$. 

\begin{figure}[t]  
\includegraphics[width=\columnwidth]{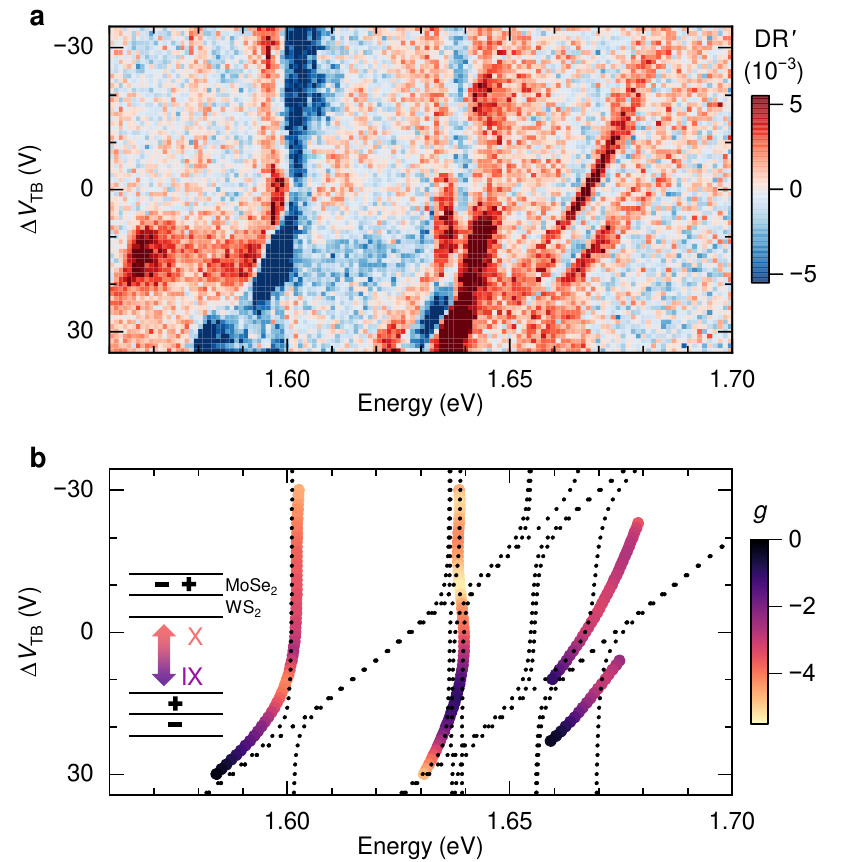}
\caption{\textbf{a}. Narrow-band modulation spectroscopy signal of MoSe$_2$/WS$_2$ as a function of the applied out-of-plane electric field. \textbf{b}. Dispersion of the eigenvalues of the Hamiltonian in Eq.~(\ref{HamMatrix}) as a function of the electric field and experimental exciton $g$-factors of the respective resonances color-coded from $-5.5$ (yellow) to $0$ (black) .}
\label{fig3}
\end{figure}

By fitting the model to electric field dependent data shown below (Figs.~\ref{fig2} and \ref{fig3}) we reproduce the multiplicity and relative oscillator strengths of the three moir\'e excitons with very good agreement. In Fig.~\ref{fig1}b we show the theoretical $\chi''$ obtained for the material parameters $a_\text{MoSe$_2$} = 0.3288$~nm, $a_\text{WS$_2$} = 0.3154$~nm, $M_\text{X} = 1.44m_0$, and $M_\text{IX} = 0.86m_0$~\cite{Kormanyos2015,Goryca2019} with electron mass $m_0$, and with free fitting parameters $E_\text{X} = 1617$~meV, $E_\text{IX} = 1614$~meV, $\theta = 178.8^\circ$, $|V| = 9.3$~meV, $\text{arg}(V) = \pi/5$, $t = 3$~meV and $\gamma_0 = 4$~meV. We emphasize that although $E_\text{IX} < E_\text{X}$, the fitting implies a type~I heterostructure due to the relation $E_\text{IX} > \min[E_\text{X} + V (\mathbf{r})] \approx 1566$~meV, with a CB offset in the order of 50~meV. The remaining fit parameters for a given twist angle ($|V|$, $\text{arg}(V)$, and $t$) determine the layer character of moir\'e excitons as well as their energy separations and oscillator strengths.

The validity of the model is best seen by comparison with the data in Fig.~\ref{fig2}a showing the evolution of DR with an out-of-plane electric field proportional to $\Delta V_\text{TB}$. As already pointed out in previous studies~\cite{Tang2021,Tang2022}, the peaks $M_1$ and $M_2$ behave as intralayer states with vanishingly small linear slopes of the first-order Stark effect. For these peaks, this implies that they can be captured in terms of the non-resonant moir\'e potential $V$, whereas interlayer tunneling plays a negligible role. The peak $M_3$, on the contrary, exhibits two branches with finite dispersion indicative of an anticrossing of X and IX states with weak coupling, which necessitates a non-vanishing hopping parameter $t$ in the model. Also, $M_1$ and $M_2$ show minor narrowing with a small red-shift at high $\Delta V_\text{TB} \geq 18$V, indicating the presence of dark IX resonances. 

To describe this behavior we compute the evolution of $\chi'' (\omega)$ as a function of $\Delta V_\text{TB}$ by shifting the IX resonance energies in (\ref{IX}) as 
\begin{equation}
  \mathcal{E}(\Delta V_\text{TB}) = \mathcal{E} - \frac{e d \Delta V_\text{TB}}{\varepsilon l},
  \label{field}
\end{equation}
where $e$ is the electron charge, $d=0.6$~nm the distance between the two TMD layers, $l=110$~nm the distance between the gates, and $\varepsilon \approx 4$~\cite{Mak_Shan_WSe2_IX_2018}. The resulting susceptibility, plotted in Fig.~\ref{fig2}b, captures the main features of the experimental data such as the relative strengths of the peaks, the strong intralayer character of $M_1$ and $M_2$, the anticrossing of the $M_3$ doublet and the IX perturbation of $M_1$ at high positive fields. At the same time, it predicts a weak coupling of $M_2$ with an IX state at $\Delta V_\text{TB} \approx -10$~V which can not be observed in white-light DR. 
%
%
\begin{figure*}[t]  
\includegraphics{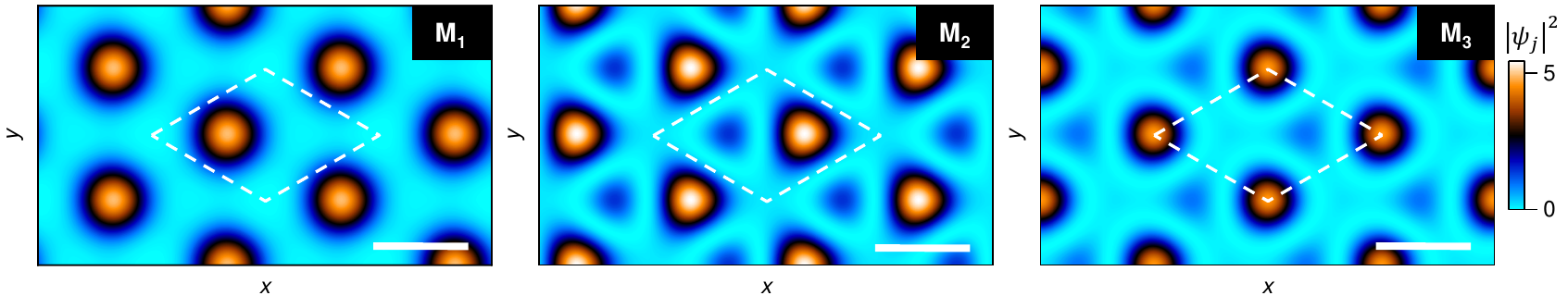}
\caption{Spatial distribution of the exciton wavefunction for the three lowest-energy bright states $M_1$, $M_2$ and $M_3$ (all scale-bars are $5$~nm). All three states are located in different positions of the moir\'e supercell delimited by the dashed lines.}
\label{fig4}
\end{figure*}

To improve the sensitivity to interlayer states we repeated the field-dependent measurement with the modulation-spectroscopy technique -- an approach that uses a narrow-band tunable laser to measure differential reflectance $\text{DR}'$~\cite{modulation2003,Hoegele2004,BarreHeinz2022}. The measurement is performed by modulating one of the gates of the device by a small ac-voltage ($V_{\mathrm{ac}}$ in Fig.~\ref{fig1}a) and using a lock-in amplifier to detect the reflected signal at the same frequency, $R_\text{ac}$, simultaneously with the dc part of the photosignal, $R_\text{dc}$~\cite{SuppMat}. In Fig.~\ref{fig3}a we show $\text{DR}' = R_\text{ac}  /R_\text{dc}$ with signatures of both intra- and interlayer states: the interlayer character of the $M_3$ doublet becomes much more prominent compared to the white-light DR data, the red-shifts of $M_1$ and $M_2$ at high $\Delta V_\text{TB}$ confirm the admixing of IX excitons, and the contrast change of $M_2$ at $\Delta V_\text{TB} \approx -20$~V suggests the coupling to a dark IX state made visible with this technique.

In Fig.~\ref{fig3}b we show the dispersion of the peaks extracted from Fig.~\ref{fig3}a alongside the evolution of the eigenvalues of the Hamiltonian in Eq.~(\ref{HamMatrix}) as a function of $\Delta V_\text{TB}$. In contrast to Fig.~\ref{fig2}b, where the oscillator strengths of the three bright excitons are visible, here the dotted lines indicate the eigenvalues corresponding to all 13 moir\'e exciton bands irrespective of their oscillator strength. The experimental dispersion of both $M_1$ and $M_2$ in Fig.~\ref{fig3}a is well reproduced, and the strong anticrossing of the $M_3$ doublet~\cite{SuppMat} is qualitatively captured within our theoretical model. We note that although the magnitude of the X-IX coupling changes between the $M_2$ and $M_3$ resonances, they are both controlled by the same parameter $t$, and introducing different hopping parameters for different mBZs~\cite{Tang2021} would allow to improve on the quantitative agreement with the data. Finally, we find that in the neutral regime the first IX state -- which is momentum-dark and can not be observed in Fig.~\ref{fig2} -- lies $30$~meV above the ground state $M_1$, indicating type~I character for the studied MoSe$_2$/WS$_2$ heterostack.

Additionally to the electric field dependence we studied the Zeeman effect of moir\'e excitons by repeating the modulation-spectroscopy measurements for out-of-plane magnetic fields in the range of $B=\pm 6T$~\cite{SuppMat}. The colors of the data in Fig.~\ref{fig3}b show the $g$-factors of the respective peaks for different electric fields, ranging from $g=-5.5$ (yellow) through $-2$ (purple) to $0$ (black). We observe that for electric fields where $M_1$ and $M_2$ follow vertical lines of zero Stark effect due to nearly-pure MoSe$_2$ character, the corresponding $g$-factors are close to the fundamental MoSe$_2$ A-exciton with $g_A \approx -4$. Near the IX-X anticrossings, on the contrary, all three states exhibit sizable changes in the $g$-factors reaching values up to $0$. The effect is most pronounced for the dispersive branch of $M_1$ at high positive electric fields as well as the $M_3$ doublet, which we attribute to field-induced hybridization with interlayer states. The values of the exciton $g$-factors depend on both the degree of layer hybridization and the exciton momentum~\cite{Paulo_2023, Gobato_Gfactors2022}, making a full quantitative description of this behavior out of scope for this work. The overall trend, however, is consistent with the intralayer character of the peaks $M_1$ and $M_2$ at negative electric fields, and with interlayer admixing near anticrossings captured by our theoretical model.

Finally, for zero electric and magnetic fields, we illustrate the difference of the three bright moir\'e excitons by plotting the spatial distributions of their wavefunctions in Fig.~\ref{fig4}~\cite{SuppMat}. Remarkably, all three states are located at different points of the moir\'e supercell and exhibit different spatial distributions: the ground state exciton $M_1$ is tightly localized, whereas both $M_2$ and $M_3$ have non-negligible spatial extents. We note that a detailed modeling of such behavior requires the explicit knowledge of both the electron and the hole potential~\cite{naik_intralayer_2022}, whereas in our model, the spatial characteristics of moir\'e excitons are captured entirely through the exciton potential $V$.

In summary, our work demonstrates that due to the close CB alignment between MoSe$_2$ and WS$_2$, full description of moir\'e exciton phenomena requires exciton mixing from a non-resonant moir\'e potential as well as resonant interlayer hybridization. We provide a simple model that highlights the importance of both effects and accurately captures the main experimental observations, and which general applicability will open avenues for future studies in related two-dimensional moir\'e structures with significance of both intra- and interlayer effects.

\begin{acknowledgments}
We acknowledge funding by the European Research Council (ERC) under the Grant Agreement No.~772195 as well as the Deutsche Forschungsgemeinschaft (DFG, German Research Foundation) within the Priority Programme SPP~2244 2DMP and the Germany's Excellence Strategy EXC-2111-390814868. B.\,P acknowledges funding by IMPRS-QST. S.\,M. and I.\,B. acknowledge support from the Alexander von Humboldt Foundation. X.\,H. and A.\,S.\,B. received funding from the European Union's Framework Programme for Research and Innovation Horizon 2020 (2014--2020) under the Marie Sk{\l}odowska-Curie Grant Agreement No.~754388 (LMUResearchFellows) and from LMUexcellent, funded by the Federal Ministry of Education and Research (BMBF) and the Free State of Bavaria under the Excellence Strategy of the German Federal Government and the L{\"a}nder. Z.\,L. was supported by the China Scholarship Council (CSC), No. 201808140196. K.W. and T.T. acknowledge support from JSPS KAKENHI (Grant Numbers 19H05790, 20H00354 and 21H05233).
\end{acknowledgments}
\vspace{11pt}
\noindent B.\,P., J.\,S., and S.\,M. contributed equally to this work.


\begin{thebibliography}{45}%
\makeatletter
\providecommand \@ifxundefined [1]{%
 \@ifx{#1\undefined}
}%
\providecommand \@ifnum [1]{%
 \ifnum #1\expandafter \@firstoftwo
 \else \expandafter \@secondoftwo
 \fi
}%
\providecommand \@ifx [1]{%
 \ifx #1\expandafter \@firstoftwo
 \else \expandafter \@secondoftwo
 \fi
}%
\providecommand \natexlab [1]{#1}%
\providecommand \enquote  [1]{``#1''}%
\providecommand \bibnamefont  [1]{#1}%
\providecommand \bibfnamefont [1]{#1}%
\providecommand \citenamefont [1]{#1}%
\providecommand \href@noop [0]{\@secondoftwo}%
\providecommand \href [0]{\begingroup \@sanitize@url \@href}%
\providecommand \@href[1]{\@@startlink{#1}\@@href}%
\providecommand \@@href[1]{\endgroup#1\@@endlink}%
\providecommand \@sanitize@url [0]{\catcode `\\12\catcode `\$12\catcode
  `\&12\catcode `\#12\catcode `\^12\catcode `\_12\catcode `\%12\relax}%
\providecommand \@@startlink[1]{}%
\providecommand \@@endlink[0]{}%
\providecommand \url  [0]{\begingroup\@sanitize@url \@url }%
\providecommand \@url [1]{\endgroup\@href {#1}{\urlprefix }}%
\providecommand \urlprefix  [0]{URL }%
\providecommand \Eprint [0]{\href }%
\providecommand \doibase [0]{https://doi.org/}%
\providecommand \selectlanguage [0]{\@gobble}%
\providecommand \bibinfo  [0]{\@secondoftwo}%
\providecommand \bibfield  [0]{\@secondoftwo}%
\providecommand \translation [1]{[#1]}%
\providecommand \BibitemOpen [0]{}%
\providecommand \bibitemStop [0]{}%
\providecommand \bibitemNoStop [0]{.\EOS\space}%
\providecommand \EOS [0]{\spacefactor3000\relax}%
\providecommand \BibitemShut  [1]{\csname bibitem#1\endcsname}%
\let\auto@bib@innerbib\@empty
\bibitem [{\citenamefont {Wang}\ \emph
  {et~al.}(2018{\natexlab{a}})\citenamefont {Wang}, \citenamefont {Chernikov},
  \citenamefont {Glazov}, \citenamefont {Heinz}, \citenamefont {Marie},
  \citenamefont {Amand},\ and\ \citenamefont {Urbaszek}}]{Colloquium_Wang2018}%
  \BibitemOpen
  \bibfield  {author} {\bibinfo {author} {\bibfnamefont {G.}~\bibnamefont
  {Wang}}, \bibinfo {author} {\bibfnamefont {A.}~\bibnamefont {Chernikov}},
  \bibinfo {author} {\bibfnamefont {M.~M.}\ \bibnamefont {Glazov}}, \bibinfo
  {author} {\bibfnamefont {T.~F.}\ \bibnamefont {Heinz}}, \bibinfo {author}
  {\bibfnamefont {X.}~\bibnamefont {Marie}}, \bibinfo {author} {\bibfnamefont
  {T.}~\bibnamefont {Amand}},\ and\ \bibinfo {author} {\bibfnamefont
  {B.}~\bibnamefont {Urbaszek}},\ }\bibfield  {title} {\bibinfo {title}
  {Colloquium: Excitons in atomically thin transition metal dichalcogenides},\
  }\href {https://doi.org/10.1103/RevModPhys.90.021001} {\bibfield  {journal}
  {\bibinfo  {journal} {Rev. Mod. Phys.}\ }\textbf {\bibinfo {volume} {90}},\
  \bibinfo {pages} {021001} (\bibinfo {year} {2018}{\natexlab{a}})}\BibitemShut
  {NoStop}%
\bibitem [{\citenamefont {Koperski}\ \emph {et~al.}(2017)\citenamefont
  {Koperski}, \citenamefont {Molas}, \citenamefont {Arora}, \citenamefont
  {Nogajewski}, \citenamefont {Slobodeniuk}, \citenamefont {Faugeras},\ and\
  \citenamefont {Potemski}}]{FaugerasPotemski2017}%
  \BibitemOpen
  \bibfield  {author} {\bibinfo {author} {\bibfnamefont {M.}~\bibnamefont
  {Koperski}}, \bibinfo {author} {\bibfnamefont {M.~R.}\ \bibnamefont {Molas}},
  \bibinfo {author} {\bibfnamefont {A.}~\bibnamefont {Arora}}, \bibinfo
  {author} {\bibfnamefont {K.}~\bibnamefont {Nogajewski}}, \bibinfo {author}
  {\bibfnamefont {A.~O.}\ \bibnamefont {Slobodeniuk}}, \bibinfo {author}
  {\bibfnamefont {C.}~\bibnamefont {Faugeras}},\ and\ \bibinfo {author}
  {\bibfnamefont {M.}~\bibnamefont {Potemski}},\ }\bibfield  {title} {\bibinfo
  {title} {Optical properties of atomically thin transition metal
  dichalcogenides: observations and puzzles},\ }\href
  {https://doi.org/10.1515/nanoph-2016-0165} {\bibfield  {journal} {\bibinfo
  {journal} {Nanophotonics}\ }\textbf {\bibinfo {volume} {6}},\ \bibinfo
  {pages} {1289} (\bibinfo {year} {2017})}\BibitemShut {NoStop}%
\bibitem [{\citenamefont {Manzeli}\ \emph {et~al.}(2017)\citenamefont
  {Manzeli}, \citenamefont {Ovchinnikov}, \citenamefont {Pasquier},
  \citenamefont {Yazyev},\ and\ \citenamefont {Kis}}]{manzeli_2d_2017}%
  \BibitemOpen
  \bibfield  {author} {\bibinfo {author} {\bibfnamefont {S.}~\bibnamefont
  {Manzeli}}, \bibinfo {author} {\bibfnamefont {D.}~\bibnamefont
  {Ovchinnikov}}, \bibinfo {author} {\bibfnamefont {D.}~\bibnamefont
  {Pasquier}}, \bibinfo {author} {\bibfnamefont {O.~V.}\ \bibnamefont
  {Yazyev}},\ and\ \bibinfo {author} {\bibfnamefont {A.}~\bibnamefont {Kis}},\
  }\bibfield  {title} {\bibinfo {title} {{2D transition metal
  dichalcogenides}},\ }\href {https://doi.org/10.1038/natrevmats.2017.33}
  {\bibfield  {journal} {\bibinfo  {journal} {Nat. Rev. Mater.}\ }\textbf
  {\bibinfo {volume} {2}},\ \bibinfo {pages} {1} (\bibinfo {year}
  {2017})}\BibitemShut {NoStop}%
\bibitem [{\citenamefont {Geim}\ and\ \citenamefont
  {Grigorieva}(2013)}]{Geim2013}%
  \BibitemOpen
  \bibfield  {author} {\bibinfo {author} {\bibfnamefont {A.~K.}\ \bibnamefont
  {Geim}}\ and\ \bibinfo {author} {\bibfnamefont {I.~V.}\ \bibnamefont
  {Grigorieva}},\ }\bibfield  {title} {\bibinfo {title} {Van der {W}aals
  heterostructures},\ }\href {https://doi.org/10.1038/nature12385} {\bibfield
  {journal} {\bibinfo  {journal} {Nature}\ }\textbf {\bibinfo {volume} {499}},\
  \bibinfo {pages} {419} (\bibinfo {year} {2013})}\BibitemShut {NoStop}%
\bibitem [{\citenamefont {Balents}\ \emph {et~al.}(2020)\citenamefont
  {Balents}, \citenamefont {Dean}, \citenamefont {Efetov},\ and\ \citenamefont
  {Young}}]{superconductivity_flat_bands_2020}%
  \BibitemOpen
  \bibfield  {author} {\bibinfo {author} {\bibfnamefont {L.}~\bibnamefont
  {Balents}}, \bibinfo {author} {\bibfnamefont {C.~R.}\ \bibnamefont {Dean}},
  \bibinfo {author} {\bibfnamefont {D.~K.}\ \bibnamefont {Efetov}},\ and\
  \bibinfo {author} {\bibfnamefont {A.~F.}\ \bibnamefont {Young}},\ }\bibfield
  {title} {\bibinfo {title} {{Superconductivity and strong correlations in
  moir{\ifmmode\acute{e}\else\'{e}\fi} flat bands}},\ }\href
  {https://doi.org/10.1038/s41567-020-0906-9} {\bibfield  {journal} {\bibinfo
  {journal} {Nat. Phys.}\ }\textbf {\bibinfo {volume} {16}},\ \bibinfo {pages}
  {725} (\bibinfo {year} {2020})}\BibitemShut {NoStop}%
\bibitem [{\citenamefont {Bistritzer}\ and\ \citenamefont
  {MacDonald}(2011)}]{Bistritzer2011}%
  \BibitemOpen
  \bibfield  {author} {\bibinfo {author} {\bibfnamefont {R.}~\bibnamefont
  {Bistritzer}}\ and\ \bibinfo {author} {\bibfnamefont {A.~H.}\ \bibnamefont
  {MacDonald}},\ }\bibfield  {title} {\bibinfo {title}
  {{Moir{\ifmmode\acute{e}\else\'{e}\fi} bands in twisted double-layer
  graphene}},\ }\href {https://doi.org/10.1073/pnas.1108174108} {\bibfield
  {journal} {\bibinfo  {journal} {Proc. Natl. Acad. Sci. U.S.A.}\ }\textbf
  {\bibinfo {volume} {108}},\ \bibinfo {pages} {12233} (\bibinfo {year}
  {2011})}\BibitemShut {NoStop}%
\bibitem [{\citenamefont {Mak}\ and\ \citenamefont
  {Shan}(2022)}]{mak_semiconductor_2022}%
  \BibitemOpen
  \bibfield  {author} {\bibinfo {author} {\bibfnamefont {K.~F.}\ \bibnamefont
  {Mak}}\ and\ \bibinfo {author} {\bibfnamefont {J.}~\bibnamefont {Shan}},\
  }\bibfield  {title} {\bibinfo {title} {{Semiconductor
  moir{\ifmmode\acute{e}\else\'{e}\fi} materials}},\ }\href
  {https://doi.org/10.1038/s41565-022-01165-6} {\bibfield  {journal} {\bibinfo
  {journal} {Nat. Nanotechnol.}\ }\textbf {\bibinfo {volume} {17}},\ \bibinfo
  {pages} {686} (\bibinfo {year} {2022})}\BibitemShut {NoStop}%
\bibitem [{\citenamefont {Hermann}(2012)}]{Hermann2012}%
  \BibitemOpen
  \bibfield  {author} {\bibinfo {author} {\bibfnamefont {K.}~\bibnamefont
  {Hermann}},\ }\bibfield  {title} {\bibinfo {title} {{Periodic overlayers and
  moir{\ifmmode\acute{e}\else\'{e}\fi} patterns: theoretical studies of
  geometric properties}},\ }\href
  {https://doi.org/10.1088/0953-8984/24/31/314210} {\bibfield  {journal}
  {\bibinfo  {journal} {J. Phys.: Condens. Matter}\ }\textbf {\bibinfo {volume}
  {24}},\ \bibinfo {pages} {314210} (\bibinfo {year} {2012})}\BibitemShut
  {NoStop}%
\bibitem [{\citenamefont {Ribeiro-Palau}\ \emph {et~al.}(2018)\citenamefont
  {Ribeiro-Palau}, \citenamefont {Zhang}, \citenamefont {Watanabe},
  \citenamefont {Taniguchi}, \citenamefont {Hone},\ and\ \citenamefont
  {Dean}}]{Dean2018_Twistable_Graphene}%
  \BibitemOpen
  \bibfield  {author} {\bibinfo {author} {\bibfnamefont {R.}~\bibnamefont
  {Ribeiro-Palau}}, \bibinfo {author} {\bibfnamefont {C.}~\bibnamefont
  {Zhang}}, \bibinfo {author} {\bibfnamefont {K.}~\bibnamefont {Watanabe}},
  \bibinfo {author} {\bibfnamefont {T.}~\bibnamefont {Taniguchi}}, \bibinfo
  {author} {\bibfnamefont {J.}~\bibnamefont {Hone}},\ and\ \bibinfo {author}
  {\bibfnamefont {C.~R.}\ \bibnamefont {Dean}},\ }\bibfield  {title} {\bibinfo
  {title} {{Twistable electronics with dynamically rotatable
  heterostructures}},\ }\href {https://doi.org/10.1126/science.aat6981}
  {\bibfield  {journal} {\bibinfo  {journal} {Science}\ }\textbf {\bibinfo
  {volume} {361}},\ \bibinfo {pages} {690} (\bibinfo {year}
  {2018})}\BibitemShut {NoStop}%
\bibitem [{\citenamefont {Wilson}\ \emph {et~al.}(2021)\citenamefont {Wilson},
  \citenamefont {Yao}, \citenamefont {Shan},\ and\ \citenamefont
  {Xu}}]{Wilson2021}%
  \BibitemOpen
  \bibfield  {author} {\bibinfo {author} {\bibfnamefont {N.~P.}\ \bibnamefont
  {Wilson}}, \bibinfo {author} {\bibfnamefont {W.}~\bibnamefont {Yao}},
  \bibinfo {author} {\bibfnamefont {J.}~\bibnamefont {Shan}},\ and\ \bibinfo
  {author} {\bibfnamefont {X.}~\bibnamefont {Xu}},\ }\bibfield  {title}
  {\bibinfo {title} {{Excitons and emergent quantum phenomena in stacked 2D
  semiconductors}},\ }\href {https://doi.org/10.1038/s41586-021-03979-1}
  {\bibfield  {journal} {\bibinfo  {journal} {Nature}\ }\textbf {\bibinfo
  {volume} {599}},\ \bibinfo {pages} {383} (\bibinfo {year}
  {2021})}\BibitemShut {NoStop}%
\bibitem [{\citenamefont {Huang}\ \emph {et~al.}(2022)\citenamefont {Huang},
  \citenamefont {Choi}, \citenamefont {Shih},\ and\ \citenamefont
  {Li}}]{huang_excitons_2022}%
  \BibitemOpen
  \bibfield  {author} {\bibinfo {author} {\bibfnamefont {D.}~\bibnamefont
  {Huang}}, \bibinfo {author} {\bibfnamefont {J.}~\bibnamefont {Choi}},
  \bibinfo {author} {\bibfnamefont {C.-K.}\ \bibnamefont {Shih}},\ and\
  \bibinfo {author} {\bibfnamefont {X.}~\bibnamefont {Li}},\ }\bibfield
  {title} {\bibinfo {title} {{Excitons in semiconductor
  moir{\ifmmode\acute{e}\else\'{e}\fi} superlattices}},\ }\href
  {https://doi.org/10.1038/s41565-021-01068-y} {\bibfield  {journal} {\bibinfo
  {journal} {Nat. Nanotechnol.}\ }\textbf {\bibinfo {volume} {17}},\ \bibinfo
  {pages} {227} (\bibinfo {year} {2022})}\BibitemShut {NoStop}%
\bibitem [{\citenamefont {Tang}\ \emph {et~al.}(2020)\citenamefont {Tang},
  \citenamefont {Li}, \citenamefont {Li}, \citenamefont {Xu}, \citenamefont
  {Liu}, \citenamefont {Barmak}, \citenamefont {Watanabe}, \citenamefont
  {Taniguchi}, \citenamefont {MacDonald}, \citenamefont {Shan},\ and\
  \citenamefont {Mak}}]{Tang2020}%
  \BibitemOpen
  \bibfield  {author} {\bibinfo {author} {\bibfnamefont {Y.}~\bibnamefont
  {Tang}}, \bibinfo {author} {\bibfnamefont {L.}~\bibnamefont {Li}}, \bibinfo
  {author} {\bibfnamefont {T.}~\bibnamefont {Li}}, \bibinfo {author}
  {\bibfnamefont {Y.}~\bibnamefont {Xu}}, \bibinfo {author} {\bibfnamefont
  {S.}~\bibnamefont {Liu}}, \bibinfo {author} {\bibfnamefont {K.}~\bibnamefont
  {Barmak}}, \bibinfo {author} {\bibfnamefont {K.}~\bibnamefont {Watanabe}},
  \bibinfo {author} {\bibfnamefont {T.}~\bibnamefont {Taniguchi}}, \bibinfo
  {author} {\bibfnamefont {A.~H.}\ \bibnamefont {MacDonald}}, \bibinfo {author}
  {\bibfnamefont {J.}~\bibnamefont {Shan}},\ and\ \bibinfo {author}
  {\bibfnamefont {K.~F.}\ \bibnamefont {Mak}},\ }\bibfield  {title} {\bibinfo
  {title} {Simulation of {H}ubbard model physics in {WSe}$_2$/{WS}$_2$ moiré
  superlattices},\ }\href {https://doi.org/10.1038/s41586-020-2085-3}
  {\bibfield  {journal} {\bibinfo  {journal} {Nature}\ }\textbf {\bibinfo
  {volume} {579}},\ \bibinfo {pages} {353} (\bibinfo {year}
  {2020})}\BibitemShut {NoStop}%
\bibitem [{\citenamefont {Campbell}\ \emph {et~al.}(2022)\citenamefont
  {Campbell}, \citenamefont {Brotons-Gisbert}, \citenamefont {Baek},
  \citenamefont {Vitale}, \citenamefont {Taniguchi}, \citenamefont {Watanabe},
  \citenamefont {Lischner},\ and\ \citenamefont
  {Gerardot}}]{Gerardot_exciton-polarons_2022}%
  \BibitemOpen
  \bibfield  {author} {\bibinfo {author} {\bibfnamefont {A.~J.}\ \bibnamefont
  {Campbell}}, \bibinfo {author} {\bibfnamefont {M.}~\bibnamefont
  {Brotons-Gisbert}}, \bibinfo {author} {\bibfnamefont {H.}~\bibnamefont
  {Baek}}, \bibinfo {author} {\bibfnamefont {V.}~\bibnamefont {Vitale}},
  \bibinfo {author} {\bibfnamefont {T.}~\bibnamefont {Taniguchi}}, \bibinfo
  {author} {\bibfnamefont {K.}~\bibnamefont {Watanabe}}, \bibinfo {author}
  {\bibfnamefont {J.}~\bibnamefont {Lischner}},\ and\ \bibinfo {author}
  {\bibfnamefont {B.~D.}\ \bibnamefont {Gerardot}},\ }\bibfield  {title}
  {\bibinfo {title} {{Exciton-polarons in the presence of strongly correlated
  electronic states in a MoSe2/WSe2 moir{\ifmmode\acute{e}\else\'{e}\fi}
  superlattice}},\ }\href {https://doi.org/10.1038/s41699-022-00358-w}
  {\bibfield  {journal} {\bibinfo  {journal} {npj 2D Mater. Appl.}\ }\textbf
  {\bibinfo {volume} {6}},\ \bibinfo {pages} {1} (\bibinfo {year}
  {2022})}\BibitemShut {NoStop}%
\bibitem [{\citenamefont {Regan}\ \emph {et~al.}(2020)\citenamefont {Regan},
  \citenamefont {Wang}, \citenamefont {Jin}, \citenamefont {Bakti~Utama},
  \citenamefont {Gao}, \citenamefont {Wei}, \citenamefont {Zhao}, \citenamefont
  {Zhao}, \citenamefont {Zhang}, \citenamefont {Yumigeta}, \citenamefont
  {Blei}, \citenamefont {Carlström}, \citenamefont {Watanabe}, \citenamefont
  {Taniguchi}, \citenamefont {Tongay}, \citenamefont {Crommie}, \citenamefont
  {Zettl},\ and\ \citenamefont {Wang}}]{Regan2020}%
  \BibitemOpen
  \bibfield  {author} {\bibinfo {author} {\bibfnamefont {E.~C.}\ \bibnamefont
  {Regan}}, \bibinfo {author} {\bibfnamefont {D.}~\bibnamefont {Wang}},
  \bibinfo {author} {\bibfnamefont {C.}~\bibnamefont {Jin}}, \bibinfo {author}
  {\bibfnamefont {M.~I.}\ \bibnamefont {Bakti~Utama}}, \bibinfo {author}
  {\bibfnamefont {B.}~\bibnamefont {Gao}}, \bibinfo {author} {\bibfnamefont
  {X.}~\bibnamefont {Wei}}, \bibinfo {author} {\bibfnamefont {S.}~\bibnamefont
  {Zhao}}, \bibinfo {author} {\bibfnamefont {W.}~\bibnamefont {Zhao}}, \bibinfo
  {author} {\bibfnamefont {Z.}~\bibnamefont {Zhang}}, \bibinfo {author}
  {\bibfnamefont {K.}~\bibnamefont {Yumigeta}}, \bibinfo {author}
  {\bibfnamefont {M.}~\bibnamefont {Blei}}, \bibinfo {author} {\bibfnamefont
  {J.~D.}\ \bibnamefont {Carlström}}, \bibinfo {author} {\bibfnamefont
  {K.}~\bibnamefont {Watanabe}}, \bibinfo {author} {\bibfnamefont
  {T.}~\bibnamefont {Taniguchi}}, \bibinfo {author} {\bibfnamefont
  {S.}~\bibnamefont {Tongay}}, \bibinfo {author} {\bibfnamefont
  {M.}~\bibnamefont {Crommie}}, \bibinfo {author} {\bibfnamefont
  {A.}~\bibnamefont {Zettl}},\ and\ \bibinfo {author} {\bibfnamefont
  {F.}~\bibnamefont {Wang}},\ }\bibfield  {title} {\bibinfo {title} {Mott and
  generalized {W}igner crystal states in {WSe}$_2$/{WS}$_2$ moiré
  superlattices},\ }\href {https://doi.org/10.1038/s41586-020-2092-4}
  {\bibfield  {journal} {\bibinfo  {journal} {Nature}\ }\textbf {\bibinfo
  {volume} {579}},\ \bibinfo {pages} {359} (\bibinfo {year}
  {2020})}\BibitemShut {NoStop}%
\bibitem [{\citenamefont {Zhou}\ \emph {et~al.}(2021)\citenamefont {Zhou},
  \citenamefont {Sung}, \citenamefont {Brutschea}, \citenamefont {Esterlis},
  \citenamefont {Wang}, \citenamefont {Scuri}, \citenamefont {Gelly},
  \citenamefont {Heo}, \citenamefont {Taniguchi}, \citenamefont {Watanabe},
  \citenamefont {Zaránd}, \citenamefont {Lukin}, \citenamefont {Kim},
  \citenamefont {Demler},\ and\ \citenamefont {Park}}]{Zhou2021}%
  \BibitemOpen
  \bibfield  {author} {\bibinfo {author} {\bibfnamefont {Y.}~\bibnamefont
  {Zhou}}, \bibinfo {author} {\bibfnamefont {J.}~\bibnamefont {Sung}}, \bibinfo
  {author} {\bibfnamefont {E.}~\bibnamefont {Brutschea}}, \bibinfo {author}
  {\bibfnamefont {I.}~\bibnamefont {Esterlis}}, \bibinfo {author}
  {\bibfnamefont {Y.}~\bibnamefont {Wang}}, \bibinfo {author} {\bibfnamefont
  {G.}~\bibnamefont {Scuri}}, \bibinfo {author} {\bibfnamefont {R.~J.}\
  \bibnamefont {Gelly}}, \bibinfo {author} {\bibfnamefont {H.}~\bibnamefont
  {Heo}}, \bibinfo {author} {\bibfnamefont {T.}~\bibnamefont {Taniguchi}},
  \bibinfo {author} {\bibfnamefont {K.}~\bibnamefont {Watanabe}}, \bibinfo
  {author} {\bibfnamefont {G.}~\bibnamefont {Zaránd}}, \bibinfo {author}
  {\bibfnamefont {M.~D.}\ \bibnamefont {Lukin}}, \bibinfo {author}
  {\bibfnamefont {P.}~\bibnamefont {Kim}}, \bibinfo {author} {\bibfnamefont
  {E.}~\bibnamefont {Demler}},\ and\ \bibinfo {author} {\bibfnamefont
  {H.}~\bibnamefont {Park}},\ }\bibfield  {title} {\bibinfo {title} {Bilayer
  wigner crystals in a transition metal dichalcogenide heterostructure},\
  }\href {https://doi.org/10.1038/s41586-021-03560-w} {\bibfield  {journal}
  {\bibinfo  {journal} {Nature}\ }\textbf {\bibinfo {volume} {595}},\ \bibinfo
  {pages} {48} (\bibinfo {year} {2021})}\BibitemShut {NoStop}%
\bibitem [{\citenamefont {Shimazaki}\ \emph {et~al.}(2020)\citenamefont
  {Shimazaki}, \citenamefont {Schwartz}, \citenamefont {Watanabe},
  \citenamefont {Taniguchi}, \citenamefont {Kroner},\ and\ \citenamefont
  {Imamoğlu}}]{Shimazaki2020}%
  \BibitemOpen
  \bibfield  {author} {\bibinfo {author} {\bibfnamefont {Y.}~\bibnamefont
  {Shimazaki}}, \bibinfo {author} {\bibfnamefont {I.}~\bibnamefont {Schwartz}},
  \bibinfo {author} {\bibfnamefont {K.}~\bibnamefont {Watanabe}}, \bibinfo
  {author} {\bibfnamefont {T.}~\bibnamefont {Taniguchi}}, \bibinfo {author}
  {\bibfnamefont {M.}~\bibnamefont {Kroner}},\ and\ \bibinfo {author}
  {\bibfnamefont {A.}~\bibnamefont {Imamoğlu}},\ }\bibfield  {title} {\bibinfo
  {title} {Strongly correlated electrons and hybrid excitons in a moiré
  heterostructure},\ }\href {https://doi.org/10.1038/s41586-020-2191-2}
  {\bibfield  {journal} {\bibinfo  {journal} {Nature}\ }\textbf {\bibinfo
  {volume} {580}},\ \bibinfo {pages} {472} (\bibinfo {year}
  {2020})}\BibitemShut {NoStop}%
\bibitem [{\citenamefont {Wang}\ \emph {et~al.}(2020)\citenamefont {Wang},
  \citenamefont {Shih}, \citenamefont {Ghiotto}, \citenamefont {Xian},
  \citenamefont {Rhodes}, \citenamefont {Tan}, \citenamefont {Claassen},
  \citenamefont {Kennes}, \citenamefont {Bai}, \citenamefont {Kim},
  \citenamefont {Watanabe}, \citenamefont {Taniguchi}, \citenamefont {Zhu},
  \citenamefont {Hone}, \citenamefont {Rubio}, \citenamefont {Pasupathy},\ and\
  \citenamefont {Dean}}]{Wang2020a}%
  \BibitemOpen
  \bibfield  {author} {\bibinfo {author} {\bibfnamefont {L.}~\bibnamefont
  {Wang}}, \bibinfo {author} {\bibfnamefont {E.-M.}\ \bibnamefont {Shih}},
  \bibinfo {author} {\bibfnamefont {A.}~\bibnamefont {Ghiotto}}, \bibinfo
  {author} {\bibfnamefont {L.}~\bibnamefont {Xian}}, \bibinfo {author}
  {\bibfnamefont {D.~A.}\ \bibnamefont {Rhodes}}, \bibinfo {author}
  {\bibfnamefont {C.}~\bibnamefont {Tan}}, \bibinfo {author} {\bibfnamefont
  {M.}~\bibnamefont {Claassen}}, \bibinfo {author} {\bibfnamefont {D.~M.}\
  \bibnamefont {Kennes}}, \bibinfo {author} {\bibfnamefont {Y.}~\bibnamefont
  {Bai}}, \bibinfo {author} {\bibfnamefont {B.}~\bibnamefont {Kim}}, \bibinfo
  {author} {\bibfnamefont {K.}~\bibnamefont {Watanabe}}, \bibinfo {author}
  {\bibfnamefont {T.}~\bibnamefont {Taniguchi}}, \bibinfo {author}
  {\bibfnamefont {X.}~\bibnamefont {Zhu}}, \bibinfo {author} {\bibfnamefont
  {J.}~\bibnamefont {Hone}}, \bibinfo {author} {\bibfnamefont {A.}~\bibnamefont
  {Rubio}}, \bibinfo {author} {\bibfnamefont {A.~N.}\ \bibnamefont
  {Pasupathy}},\ and\ \bibinfo {author} {\bibfnamefont {C.~R.}\ \bibnamefont
  {Dean}},\ }\bibfield  {title} {\bibinfo {title} {Correlated electronic phases
  in twisted bilayer transition metal dichalcogenides},\ }\href
  {https://doi.org/10.1038/s41563-020-0708-6} {\bibfield  {journal} {\bibinfo
  {journal} {Nat. Mater.}\ }\textbf {\bibinfo {volume} {19}},\ \bibinfo {pages}
  {861} (\bibinfo {year} {2020})}\BibitemShut {NoStop}%
\bibitem [{\citenamefont {Huang}\ \emph {et~al.}(2021)\citenamefont {Huang},
  \citenamefont {Wang}, \citenamefont {Miao}, \citenamefont {Wang},
  \citenamefont {Li}, \citenamefont {Lian}, \citenamefont {Taniguchi},
  \citenamefont {Watanabe}, \citenamefont {Okamoto}, \citenamefont {Xiao},
  \citenamefont {Shi},\ and\ \citenamefont {Cui}}]{Huang2021}%
  \BibitemOpen
  \bibfield  {author} {\bibinfo {author} {\bibfnamefont {X.}~\bibnamefont
  {Huang}}, \bibinfo {author} {\bibfnamefont {T.}~\bibnamefont {Wang}},
  \bibinfo {author} {\bibfnamefont {S.}~\bibnamefont {Miao}}, \bibinfo {author}
  {\bibfnamefont {C.}~\bibnamefont {Wang}}, \bibinfo {author} {\bibfnamefont
  {Z.}~\bibnamefont {Li}}, \bibinfo {author} {\bibfnamefont {Z.}~\bibnamefont
  {Lian}}, \bibinfo {author} {\bibfnamefont {T.}~\bibnamefont {Taniguchi}},
  \bibinfo {author} {\bibfnamefont {K.}~\bibnamefont {Watanabe}}, \bibinfo
  {author} {\bibfnamefont {S.}~\bibnamefont {Okamoto}}, \bibinfo {author}
  {\bibfnamefont {D.}~\bibnamefont {Xiao}}, \bibinfo {author} {\bibfnamefont
  {S.-F.}\ \bibnamefont {Shi}},\ and\ \bibinfo {author} {\bibfnamefont {Y.-T.}\
  \bibnamefont {Cui}},\ }\bibfield  {title} {\bibinfo {title} {{Correlated
  insulating states at fractional fillings of the WS$_2$/WSe$_2$ moir\'e
  lattice}},\ }\href {https://doi.org/10.1038/s41567-021-01171-w} {\bibfield
  {journal} {\bibinfo  {journal} {Nat. Phys.}\ }\textbf {\bibinfo {volume}
  {17}},\ \bibinfo {pages} {715} (\bibinfo {year} {2021})}\BibitemShut
  {NoStop}%
\bibitem [{\citenamefont {Ghiotto}\ \emph {et~al.}(2021)\citenamefont
  {Ghiotto}, \citenamefont {Shih}, \citenamefont {Pereira}, \citenamefont
  {Rhodes}, \citenamefont {Kim}, \citenamefont {Zang}, \citenamefont {Millis},
  \citenamefont {Watanabe}, \citenamefont {Taniguchi}, \citenamefont {Hone},
  \citenamefont {Wang}, \citenamefont {Dean},\ and\ \citenamefont
  {Pasupathy}}]{Ghiotto2021}%
  \BibitemOpen
  \bibfield  {author} {\bibinfo {author} {\bibfnamefont {A.}~\bibnamefont
  {Ghiotto}}, \bibinfo {author} {\bibfnamefont {E.-M.}\ \bibnamefont {Shih}},
  \bibinfo {author} {\bibfnamefont {G.~S. S.~G.}\ \bibnamefont {Pereira}},
  \bibinfo {author} {\bibfnamefont {D.~A.}\ \bibnamefont {Rhodes}}, \bibinfo
  {author} {\bibfnamefont {B.}~\bibnamefont {Kim}}, \bibinfo {author}
  {\bibfnamefont {J.}~\bibnamefont {Zang}}, \bibinfo {author} {\bibfnamefont
  {A.~J.}\ \bibnamefont {Millis}}, \bibinfo {author} {\bibfnamefont
  {K.}~\bibnamefont {Watanabe}}, \bibinfo {author} {\bibfnamefont
  {T.}~\bibnamefont {Taniguchi}}, \bibinfo {author} {\bibfnamefont {J.~C.}\
  \bibnamefont {Hone}}, \bibinfo {author} {\bibfnamefont {L.}~\bibnamefont
  {Wang}}, \bibinfo {author} {\bibfnamefont {C.~R.}\ \bibnamefont {Dean}},\
  and\ \bibinfo {author} {\bibfnamefont {A.~N.}\ \bibnamefont {Pasupathy}},\
  }\bibfield  {title} {\bibinfo {title} {Quantum criticality in twisted
  transition metal dichalcogenides},\ }\href
  {https://doi.org/10.1038/s41586-021-03815-6} {\bibfield  {journal} {\bibinfo
  {journal} {Nature}\ }\textbf {\bibinfo {volume} {597}},\ \bibinfo {pages}
  {345} (\bibinfo {year} {2021})}\BibitemShut {NoStop}%
\bibitem [{\citenamefont {Xu}\ \emph {et~al.}(2020)\citenamefont {Xu},
  \citenamefont {Liu}, \citenamefont {Rhodes}, \citenamefont {Watanabe},
  \citenamefont {Taniguchi}, \citenamefont {Hone}, \citenamefont {Elser},
  \citenamefont {Mak},\ and\ \citenamefont {Shan}}]{Xu2020}%
  \BibitemOpen
  \bibfield  {author} {\bibinfo {author} {\bibfnamefont {Y.}~\bibnamefont
  {Xu}}, \bibinfo {author} {\bibfnamefont {S.}~\bibnamefont {Liu}}, \bibinfo
  {author} {\bibfnamefont {D.~A.}\ \bibnamefont {Rhodes}}, \bibinfo {author}
  {\bibfnamefont {K.}~\bibnamefont {Watanabe}}, \bibinfo {author}
  {\bibfnamefont {T.}~\bibnamefont {Taniguchi}}, \bibinfo {author}
  {\bibfnamefont {J.}~\bibnamefont {Hone}}, \bibinfo {author} {\bibfnamefont
  {V.}~\bibnamefont {Elser}}, \bibinfo {author} {\bibfnamefont {K.~F.}\
  \bibnamefont {Mak}},\ and\ \bibinfo {author} {\bibfnamefont {J.}~\bibnamefont
  {Shan}},\ }\bibfield  {title} {\bibinfo {title} {Correlated insulating states
  at fractional fillings of moiré superlattices},\ }\href
  {https://doi.org/10.1038/s41586-020-2868-6} {\bibfield  {journal} {\bibinfo
  {journal} {Nature}\ }\textbf {\bibinfo {volume} {587}},\ \bibinfo {pages}
  {214} (\bibinfo {year} {2020})}\BibitemShut {NoStop}%
\bibitem [{\citenamefont {Li}\ \emph {et~al.}(2021{\natexlab{a}})\citenamefont
  {Li}, \citenamefont {Jiang}, \citenamefont {Li}, \citenamefont {Zhang},
  \citenamefont {Kang}, \citenamefont {Zhu}, \citenamefont {Watanabe},
  \citenamefont {Taniguchi}, \citenamefont {Chowdhury}, \citenamefont {Fu},
  \citenamefont {Shan},\ and\ \citenamefont {Mak}}]{LiMak2021mit}%
  \BibitemOpen
  \bibfield  {author} {\bibinfo {author} {\bibfnamefont {T.}~\bibnamefont
  {Li}}, \bibinfo {author} {\bibfnamefont {S.}~\bibnamefont {Jiang}}, \bibinfo
  {author} {\bibfnamefont {L.}~\bibnamefont {Li}}, \bibinfo {author}
  {\bibfnamefont {Y.}~\bibnamefont {Zhang}}, \bibinfo {author} {\bibfnamefont
  {K.}~\bibnamefont {Kang}}, \bibinfo {author} {\bibfnamefont {J.}~\bibnamefont
  {Zhu}}, \bibinfo {author} {\bibfnamefont {K.}~\bibnamefont {Watanabe}},
  \bibinfo {author} {\bibfnamefont {T.}~\bibnamefont {Taniguchi}}, \bibinfo
  {author} {\bibfnamefont {D.}~\bibnamefont {Chowdhury}}, \bibinfo {author}
  {\bibfnamefont {L.}~\bibnamefont {Fu}}, \bibinfo {author} {\bibfnamefont
  {J.}~\bibnamefont {Shan}},\ and\ \bibinfo {author} {\bibfnamefont {K.~F.}\
  \bibnamefont {Mak}},\ }\bibfield  {title} {\bibinfo {title} {{Continuous Mott
  transition in semiconductor moir\'e superlattices}},\ }\href
  {https://doi.org/10.1038/s41586-021-03853-0} {\bibfield  {journal} {\bibinfo
  {journal} {Nature}\ }\textbf {\bibinfo {volume} {597}},\ \bibinfo {pages}
  {350} (\bibinfo {year} {2021}{\natexlab{a}})}\BibitemShut {NoStop}%
\bibitem [{\citenamefont {Li}\ \emph {et~al.}(2021{\natexlab{b}})\citenamefont
  {Li}, \citenamefont {Zhu}, \citenamefont {Tang}, \citenamefont {Watanabe},
  \citenamefont {Taniguchi}, \citenamefont {Elser}, \citenamefont {Shan},\ and\
  \citenamefont {Mak}}]{Mak2021_capacitance_charge_order}%
  \BibitemOpen
  \bibfield  {author} {\bibinfo {author} {\bibfnamefont {T.}~\bibnamefont
  {Li}}, \bibinfo {author} {\bibfnamefont {J.}~\bibnamefont {Zhu}}, \bibinfo
  {author} {\bibfnamefont {Y.}~\bibnamefont {Tang}}, \bibinfo {author}
  {\bibfnamefont {K.}~\bibnamefont {Watanabe}}, \bibinfo {author}
  {\bibfnamefont {T.}~\bibnamefont {Taniguchi}}, \bibinfo {author}
  {\bibfnamefont {V.}~\bibnamefont {Elser}}, \bibinfo {author} {\bibfnamefont
  {J.}~\bibnamefont {Shan}},\ and\ \bibinfo {author} {\bibfnamefont {K.~F.}\
  \bibnamefont {Mak}},\ }\bibfield  {title} {\bibinfo {title}
  {Charge-order-enhanced capacitance in semiconductor moiré superlattices},\
  }\href {https://doi.org/10.1038/s41565-021-00955-8} {\bibfield  {journal}
  {\bibinfo  {journal} {Nat. Nanotechnol.}\ }\textbf {\bibinfo {volume} {16}},\
  \bibinfo {pages} {1068} (\bibinfo {year} {2021}{\natexlab{b}})}\BibitemShut
  {NoStop}%
\bibitem [{\citenamefont {Tang}\ \emph {et~al.}(2022)\citenamefont {Tang},
  \citenamefont {Gu}, \citenamefont {Liu}, \citenamefont {Watanabe},
  \citenamefont {Taniguchi}, \citenamefont {Hone}, \citenamefont {Mak},\ and\
  \citenamefont {Shan}}]{Tang2022}%
  \BibitemOpen
  \bibfield  {author} {\bibinfo {author} {\bibfnamefont {Y.}~\bibnamefont
  {Tang}}, \bibinfo {author} {\bibfnamefont {J.}~\bibnamefont {Gu}}, \bibinfo
  {author} {\bibfnamefont {S.}~\bibnamefont {Liu}}, \bibinfo {author}
  {\bibfnamefont {K.}~\bibnamefont {Watanabe}}, \bibinfo {author}
  {\bibfnamefont {T.}~\bibnamefont {Taniguchi}}, \bibinfo {author}
  {\bibfnamefont {J.~C.}\ \bibnamefont {Hone}}, \bibinfo {author}
  {\bibfnamefont {K.~F.}\ \bibnamefont {Mak}},\ and\ \bibinfo {author}
  {\bibfnamefont {J.}~\bibnamefont {Shan}},\ }\bibfield  {title} {\bibinfo
  {title} {{Dielectric catastrophe at the Wigner-Mott transition in a
  moir{\ifmmode\acute{e}\else\'{e}\fi} superlattice}},\ }\href
  {https://doi.org/10.1038/s41467-022-32037-1} {\bibfield  {journal} {\bibinfo
  {journal} {Nat. Commun.}\ }\textbf {\bibinfo {volume} {13}},\ \bibinfo
  {pages} {1} (\bibinfo {year} {2022})}\BibitemShut {NoStop}%
\bibitem [{\citenamefont {Zhang}\ \emph {et~al.}(2016)\citenamefont {Zhang},
  \citenamefont {Gong}, \citenamefont {Nie}, \citenamefont {Min}, \citenamefont
  {Liang}, \citenamefont {Oh}, \citenamefont {Zhang}, \citenamefont {Wang},
  \citenamefont {Hong}, \citenamefont {Colombo}, \citenamefont {Wallace},\ and\
  \citenamefont {Cho}}]{Zhang2016}%
  \BibitemOpen
  \bibfield  {author} {\bibinfo {author} {\bibfnamefont {C.}~\bibnamefont
  {Zhang}}, \bibinfo {author} {\bibfnamefont {C.}~\bibnamefont {Gong}},
  \bibinfo {author} {\bibfnamefont {Y.}~\bibnamefont {Nie}}, \bibinfo {author}
  {\bibfnamefont {K.-A.}\ \bibnamefont {Min}}, \bibinfo {author} {\bibfnamefont
  {C.}~\bibnamefont {Liang}}, \bibinfo {author} {\bibfnamefont {Y.~J.}\
  \bibnamefont {Oh}}, \bibinfo {author} {\bibfnamefont {H.}~\bibnamefont
  {Zhang}}, \bibinfo {author} {\bibfnamefont {W.}~\bibnamefont {Wang}},
  \bibinfo {author} {\bibfnamefont {S.}~\bibnamefont {Hong}}, \bibinfo {author}
  {\bibfnamefont {L.}~\bibnamefont {Colombo}}, \bibinfo {author} {\bibfnamefont
  {R.~M.}\ \bibnamefont {Wallace}},\ and\ \bibinfo {author} {\bibfnamefont
  {K.}~\bibnamefont {Cho}},\ }\bibfield  {title} {\bibinfo {title} {{Systematic
  study of electronic structure and band alignment of monolayer transition
  metal dichalcogenides in van der Waals heterostructures}},\ }\href
  {https://doi.org/10.1088/2053-1583/4/1/015026} {\bibfield  {journal}
  {\bibinfo  {journal} {2D Mater.}\ }\textbf {\bibinfo {volume} {4}},\ \bibinfo
  {pages} {015026} (\bibinfo {year} {2016})}\BibitemShut {NoStop}%
\bibitem [{\citenamefont {Zhang}\ \emph {et~al.}(2020)\citenamefont {Zhang},
  \citenamefont {Zhang}, \citenamefont {Wu}, \citenamefont {Wang},
  \citenamefont {Gogna}, \citenamefont {Hou}, \citenamefont {Watanabe},
  \citenamefont {Taniguchi}, \citenamefont {Kulkarni}, \citenamefont {Kuo},
  \citenamefont {Forrest},\ and\ \citenamefont {Deng}}]{Zhang_Deng_Twist2020}%
  \BibitemOpen
  \bibfield  {author} {\bibinfo {author} {\bibfnamefont {L.}~\bibnamefont
  {Zhang}}, \bibinfo {author} {\bibfnamefont {Z.}~\bibnamefont {Zhang}},
  \bibinfo {author} {\bibfnamefont {F.}~\bibnamefont {Wu}}, \bibinfo {author}
  {\bibfnamefont {D.}~\bibnamefont {Wang}}, \bibinfo {author} {\bibfnamefont
  {R.}~\bibnamefont {Gogna}}, \bibinfo {author} {\bibfnamefont
  {S.}~\bibnamefont {Hou}}, \bibinfo {author} {\bibfnamefont {K.}~\bibnamefont
  {Watanabe}}, \bibinfo {author} {\bibfnamefont {T.}~\bibnamefont {Taniguchi}},
  \bibinfo {author} {\bibfnamefont {K.}~\bibnamefont {Kulkarni}}, \bibinfo
  {author} {\bibfnamefont {T.}~\bibnamefont {Kuo}}, \bibinfo {author}
  {\bibfnamefont {S.~R.}\ \bibnamefont {Forrest}},\ and\ \bibinfo {author}
  {\bibfnamefont {H.}~\bibnamefont {Deng}},\ }\bibfield  {title} {\bibinfo
  {title} {Twist-angle dependence of moiré excitons in {WS2}/{MoSe2}
  heterobilayers},\ }\href {https://doi.org/10.1038/s41467-020-19466-6}
  {\bibfield  {journal} {\bibinfo  {journal} {Nat. Commun.}\ }\textbf {\bibinfo
  {volume} {11}},\ \bibinfo {pages} {5888} (\bibinfo {year}
  {2020})}\BibitemShut {NoStop}%
\bibitem [{\citenamefont {Bilgin}\ \emph {et~al.}(2015)\citenamefont {Bilgin},
  \citenamefont {Liu}, \citenamefont {Vargas}, \citenamefont {Winchester},
  \citenamefont {Man}, \citenamefont {Upmanyu}, \citenamefont {Dani},
  \citenamefont {Gupta}, \citenamefont {Talapatra}, \citenamefont {Mohite},\
  and\ \citenamefont {Kar}}]{bilgin2015chemical}%
  \BibitemOpen
  \bibfield  {author} {\bibinfo {author} {\bibfnamefont {I.}~\bibnamefont
  {Bilgin}}, \bibinfo {author} {\bibfnamefont {F.}~\bibnamefont {Liu}},
  \bibinfo {author} {\bibfnamefont {A.}~\bibnamefont {Vargas}}, \bibinfo
  {author} {\bibfnamefont {A.}~\bibnamefont {Winchester}}, \bibinfo {author}
  {\bibfnamefont {M.~K.}\ \bibnamefont {Man}}, \bibinfo {author} {\bibfnamefont
  {M.}~\bibnamefont {Upmanyu}}, \bibinfo {author} {\bibfnamefont {K.~M.}\
  \bibnamefont {Dani}}, \bibinfo {author} {\bibfnamefont {G.}~\bibnamefont
  {Gupta}}, \bibinfo {author} {\bibfnamefont {S.}~\bibnamefont {Talapatra}},
  \bibinfo {author} {\bibfnamefont {A.~D.}\ \bibnamefont {Mohite}},\ and\
  \bibinfo {author} {\bibfnamefont {S.}~\bibnamefont {Kar}},\ }\bibfield
  {title} {\bibinfo {title} {Chemical vapor deposition synthesized atomically
  thin molybdenum disulfide with optoelectronic-grade crystalline quality},\
  }\href {https://doi.org/10.1021/acsnano.5b02019} {\bibfield  {journal}
  {\bibinfo  {journal} {ACS Nano}\ }\textbf {\bibinfo {volume} {9}},\ \bibinfo
  {pages} {8822} (\bibinfo {year} {2015})}\BibitemShut {NoStop}%
\bibitem [{Sup()}]{SuppMat}%
  \BibitemOpen
  \href@noop {} {}\bibinfo {note} {See Supplemental Material for
  descriptions of the device layout and fabrication, the computation of
  $\chi''$ from DR data, the measurement setup, the limiting cases of the theoretical model, its dependence on individual parameters and the computation of the
  exciton wavefunctions, which includes Refs. [40, 46-48]}\BibitemShut {Stop}%
\bibitem [{\citenamefont {Alexeev}\ \emph {et~al.}(2019)\citenamefont
  {Alexeev}, \citenamefont {Ruiz-Tijerina}, \citenamefont {Danovich},
  \citenamefont {Hamer}, \citenamefont {Terry}, \citenamefont {Nayak},
  \citenamefont {Ahn}, \citenamefont {Pak}, \citenamefont {Lee}, \citenamefont
  {Sohn}, \citenamefont {Molas}, \citenamefont {Koperski}, \citenamefont
  {Watanabe}, \citenamefont {Taniguchi}, \citenamefont {Novoselov},
  \citenamefont {Gorbachev}, \citenamefont {Shin}, \citenamefont {Fal’ko},\
  and\ \citenamefont {Tartakovskii}}]{AlexeevTartakovsky2019}%
  \BibitemOpen
  \bibfield  {author} {\bibinfo {author} {\bibfnamefont {E.~M.}\ \bibnamefont
  {Alexeev}}, \bibinfo {author} {\bibfnamefont {D.~A.}\ \bibnamefont
  {Ruiz-Tijerina}}, \bibinfo {author} {\bibfnamefont {M.}~\bibnamefont
  {Danovich}}, \bibinfo {author} {\bibfnamefont {M.~J.}\ \bibnamefont {Hamer}},
  \bibinfo {author} {\bibfnamefont {D.~J.}\ \bibnamefont {Terry}}, \bibinfo
  {author} {\bibfnamefont {P.~K.}\ \bibnamefont {Nayak}}, \bibinfo {author}
  {\bibfnamefont {S.}~\bibnamefont {Ahn}}, \bibinfo {author} {\bibfnamefont
  {S.}~\bibnamefont {Pak}}, \bibinfo {author} {\bibfnamefont {J.}~\bibnamefont
  {Lee}}, \bibinfo {author} {\bibfnamefont {J.~I.}\ \bibnamefont {Sohn}},
  \bibinfo {author} {\bibfnamefont {M.~R.}\ \bibnamefont {Molas}}, \bibinfo
  {author} {\bibfnamefont {M.}~\bibnamefont {Koperski}}, \bibinfo {author}
  {\bibfnamefont {K.}~\bibnamefont {Watanabe}}, \bibinfo {author}
  {\bibfnamefont {T.}~\bibnamefont {Taniguchi}}, \bibinfo {author}
  {\bibfnamefont {K.~S.}\ \bibnamefont {Novoselov}}, \bibinfo {author}
  {\bibfnamefont {R.~V.}\ \bibnamefont {Gorbachev}}, \bibinfo {author}
  {\bibfnamefont {H.~S.}\ \bibnamefont {Shin}}, \bibinfo {author}
  {\bibfnamefont {V.~I.}\ \bibnamefont {Fal’ko}},\ and\ \bibinfo {author}
  {\bibfnamefont {A.~I.}\ \bibnamefont {Tartakovskii}},\ }\bibfield  {title}
  {\bibinfo {title} {Resonantly hybridized excitons in moiré superlattices in
  van der {W}aals heterostructures},\ }\href
  {https://doi.org/10.1038/s41586-019-0986-9} {\bibfield  {journal} {\bibinfo
  {journal} {Nature}\ }\textbf {\bibinfo {volume} {567}},\ \bibinfo {pages}
  {81} (\bibinfo {year} {2019})}\BibitemShut {NoStop}%
\bibitem [{\citenamefont {Tang}\ \emph {et~al.}(2021)\citenamefont {Tang},
  \citenamefont {Gu}, \citenamefont {Liu}, \citenamefont {Watanabe},
  \citenamefont {Taniguchi}, \citenamefont {Hone}, \citenamefont {Mak},\ and\
  \citenamefont {Shan}}]{Tang2021}%
  \BibitemOpen
  \bibfield  {author} {\bibinfo {author} {\bibfnamefont {Y.}~\bibnamefont
  {Tang}}, \bibinfo {author} {\bibfnamefont {J.}~\bibnamefont {Gu}}, \bibinfo
  {author} {\bibfnamefont {S.}~\bibnamefont {Liu}}, \bibinfo {author}
  {\bibfnamefont {K.}~\bibnamefont {Watanabe}}, \bibinfo {author}
  {\bibfnamefont {T.}~\bibnamefont {Taniguchi}}, \bibinfo {author}
  {\bibfnamefont {J.}~\bibnamefont {Hone}}, \bibinfo {author} {\bibfnamefont
  {K.~F.}\ \bibnamefont {Mak}},\ and\ \bibinfo {author} {\bibfnamefont
  {J.}~\bibnamefont {Shan}},\ }\bibfield  {title} {\bibinfo {title} {{Tuning
  layer-hybridized moir\'e excitons by the quantum-confined Stark effect}},\
  }\href {https://doi.org/10.1038/s41565-020-00783-2} {\bibfield  {journal}
  {\bibinfo  {journal} {Nat. Nanotechnol.}\ }\textbf {\bibinfo {volume} {16}},\
  \bibinfo {pages} {52} (\bibinfo {year} {2021})}\BibitemShut {NoStop}%
\bibitem [{\citenamefont {Ruiz-Tijerina}\ and\ \citenamefont
  {Fal'ko}(2019)}]{RuizFalko2019}%
  \BibitemOpen
  \bibfield  {author} {\bibinfo {author} {\bibfnamefont {D.~A.}\ \bibnamefont
  {Ruiz-Tijerina}}\ and\ \bibinfo {author} {\bibfnamefont {V.~I.}\ \bibnamefont
  {Fal'ko}},\ }\bibfield  {title} {\bibinfo {title} {Interlayer hybridization
  and moir\'e superlattice minibands for electrons and excitons in
  heterobilayers of transition-metal dichalcogenides},\ }\href
  {https://doi.org/10.1103/PhysRevB.99.125424} {\bibfield  {journal} {\bibinfo
  {journal} {Phys. Rev. B}\ }\textbf {\bibinfo {volume} {99}},\ \bibinfo
  {pages} {125424} (\bibinfo {year} {2019})}\BibitemShut {NoStop}%
\bibitem [{\citenamefont {Wang}\ \emph {et~al.}(2017)\citenamefont {Wang},
  \citenamefont {Wang}, \citenamefont {Yao}, \citenamefont {Liu},\ and\
  \citenamefont {Yu}}]{Yu_IXcoupling_2017}%
  \BibitemOpen
  \bibfield  {author} {\bibinfo {author} {\bibfnamefont {Y.}~\bibnamefont
  {Wang}}, \bibinfo {author} {\bibfnamefont {Z.}~\bibnamefont {Wang}}, \bibinfo
  {author} {\bibfnamefont {W.}~\bibnamefont {Yao}}, \bibinfo {author}
  {\bibfnamefont {G.-B.}\ \bibnamefont {Liu}},\ and\ \bibinfo {author}
  {\bibfnamefont {H.}~\bibnamefont {Yu}},\ }\bibfield  {title} {\bibinfo
  {title} {Interlayer coupling in commensurate and incommensurate bilayer
  structures of transition-metal dichalcogenides},\ }\href
  {https://doi.org/10.1103/PhysRevB.95.115429} {\bibfield  {journal} {\bibinfo
  {journal} {Phys. Rev. B}\ }\textbf {\bibinfo {volume} {95}},\ \bibinfo
  {pages} {115429} (\bibinfo {year} {2017})}\BibitemShut {NoStop}%
\bibitem [{\citenamefont {Wu}\ \emph {et~al.}(2017)\citenamefont {Wu},
  \citenamefont {Lovorn},\ and\ \citenamefont {MacDonald}}]{MacdonaldTopo2017}%
  \BibitemOpen
  \bibfield  {author} {\bibinfo {author} {\bibfnamefont {F.}~\bibnamefont
  {Wu}}, \bibinfo {author} {\bibfnamefont {T.}~\bibnamefont {Lovorn}},\ and\
  \bibinfo {author} {\bibfnamefont {A.~H.}\ \bibnamefont {MacDonald}},\
  }\bibfield  {title} {\bibinfo {title} {{Topological Exciton Bands in Moir\'e
  Heterojunctions}},\ }\href {https://doi.org/10.1103/PhysRevLett.118.147401}
  {\bibfield  {journal} {\bibinfo  {journal} {Phys. Rev. Lett.}\ }\textbf
  {\bibinfo {volume} {118}},\ \bibinfo {pages} {147401} (\bibinfo {year}
  {2017})}\BibitemShut {NoStop}%
\bibitem [{\citenamefont {Wu}\ \emph {et~al.}(2018{\natexlab{a}})\citenamefont
  {Wu}, \citenamefont {Lovorn},\ and\ \citenamefont
  {MacDonald}}]{MacdonaldIX2018}%
  \BibitemOpen
  \bibfield  {author} {\bibinfo {author} {\bibfnamefont {F.}~\bibnamefont
  {Wu}}, \bibinfo {author} {\bibfnamefont {T.}~\bibnamefont {Lovorn}},\ and\
  \bibinfo {author} {\bibfnamefont {A.~H.}\ \bibnamefont {MacDonald}},\
  }\bibfield  {title} {\bibinfo {title} {Theory of optical absorption by
  interlayer excitons in transition metal dichalcogenide heterobilayers},\
  }\href {https://doi.org/10.1103/PhysRevB.97.035306} {\bibfield  {journal}
  {\bibinfo  {journal} {Phys. Rev. B}\ }\textbf {\bibinfo {volume} {97}},\
  \bibinfo {pages} {035306} (\bibinfo {year} {2018}{\natexlab{a}})}\BibitemShut
  {NoStop}%
\bibitem [{\citenamefont {Wu}\ \emph {et~al.}(2018{\natexlab{b}})\citenamefont
  {Wu}, \citenamefont {Lovorn}, \citenamefont {Tutuc},\ and\ \citenamefont
  {MacDonald}}]{MacdonaldHubbard2018}%
  \BibitemOpen
  \bibfield  {author} {\bibinfo {author} {\bibfnamefont {F.}~\bibnamefont
  {Wu}}, \bibinfo {author} {\bibfnamefont {T.}~\bibnamefont {Lovorn}}, \bibinfo
  {author} {\bibfnamefont {E.}~\bibnamefont {Tutuc}},\ and\ \bibinfo {author}
  {\bibfnamefont {A.~H.}\ \bibnamefont {MacDonald}},\ }\bibfield  {title}
  {\bibinfo {title} {Hubbard model physics in transition metal dichalcogenide
  moir\'e bands},\ }\href {https://doi.org/10.1103/PhysRevLett.121.026402}
  {\bibfield  {journal} {\bibinfo  {journal} {Phys. Rev. Lett.}\ }\textbf
  {\bibinfo {volume} {121}},\ \bibinfo {pages} {026402} (\bibinfo {year}
  {2018}{\natexlab{b}})}\BibitemShut {NoStop}%
\bibitem [{\citenamefont {Tran}\ \emph {et~al.}(2019)\citenamefont {Tran},
  \citenamefont {Moody}, \citenamefont {Wu}, \citenamefont {Lu}, \citenamefont
  {Choi}, \citenamefont {Kim}, \citenamefont {Rai}, \citenamefont {Sanchez},
  \citenamefont {Quan}, \citenamefont {Singh}, \citenamefont {Embley},
  \citenamefont {Zepeda}, \citenamefont {Campbell}, \citenamefont {Autry},
  \citenamefont {Taniguchi}, \citenamefont {Watanabe}, \citenamefont {Lu},
  \citenamefont {Banerjee}, \citenamefont {Silverman}, \citenamefont {Kim},
  \citenamefont {Tutuc}, \citenamefont {Yang}, \citenamefont {MacDonald},\ and\
  \citenamefont {Li}}]{TranMacdonald2019}%
  \BibitemOpen
  \bibfield  {author} {\bibinfo {author} {\bibfnamefont {K.}~\bibnamefont
  {Tran}}, \bibinfo {author} {\bibfnamefont {G.}~\bibnamefont {Moody}},
  \bibinfo {author} {\bibfnamefont {F.}~\bibnamefont {Wu}}, \bibinfo {author}
  {\bibfnamefont {X.}~\bibnamefont {Lu}}, \bibinfo {author} {\bibfnamefont
  {J.}~\bibnamefont {Choi}}, \bibinfo {author} {\bibfnamefont {K.}~\bibnamefont
  {Kim}}, \bibinfo {author} {\bibfnamefont {A.}~\bibnamefont {Rai}}, \bibinfo
  {author} {\bibfnamefont {D.~A.}\ \bibnamefont {Sanchez}}, \bibinfo {author}
  {\bibfnamefont {J.}~\bibnamefont {Quan}}, \bibinfo {author} {\bibfnamefont
  {A.}~\bibnamefont {Singh}}, \bibinfo {author} {\bibfnamefont
  {J.}~\bibnamefont {Embley}}, \bibinfo {author} {\bibfnamefont
  {A.}~\bibnamefont {Zepeda}}, \bibinfo {author} {\bibfnamefont
  {M.}~\bibnamefont {Campbell}}, \bibinfo {author} {\bibfnamefont
  {T.}~\bibnamefont {Autry}}, \bibinfo {author} {\bibfnamefont
  {T.}~\bibnamefont {Taniguchi}}, \bibinfo {author} {\bibfnamefont
  {K.}~\bibnamefont {Watanabe}}, \bibinfo {author} {\bibfnamefont
  {N.}~\bibnamefont {Lu}}, \bibinfo {author} {\bibfnamefont {S.~K.}\
  \bibnamefont {Banerjee}}, \bibinfo {author} {\bibfnamefont {K.~L.}\
  \bibnamefont {Silverman}}, \bibinfo {author} {\bibfnamefont {S.}~\bibnamefont
  {Kim}}, \bibinfo {author} {\bibfnamefont {E.}~\bibnamefont {Tutuc}}, \bibinfo
  {author} {\bibfnamefont {L.}~\bibnamefont {Yang}}, \bibinfo {author}
  {\bibfnamefont {A.~H.}\ \bibnamefont {MacDonald}},\ and\ \bibinfo {author}
  {\bibfnamefont {X.}~\bibnamefont {Li}},\ }\bibfield  {title} {\bibinfo
  {title} {Evidence for moiré excitons in van der {W}aals heterostructures},\
  }\href {https://doi.org/10.1038/s41586-019-0975-z} {\bibfield  {journal}
  {\bibinfo  {journal} {Nature}\ }\textbf {\bibinfo {volume} {567}},\ \bibinfo
  {pages} {71} (\bibinfo {year} {2019})}\BibitemShut {NoStop}%
\bibitem [{\citenamefont {Jin}\ \emph {et~al.}(2019)\citenamefont {Jin},
  \citenamefont {Regan}, \citenamefont {Yan}, \citenamefont {Iqbal
  Bakti~Utama}, \citenamefont {Wang}, \citenamefont {Zhao}, \citenamefont
  {Qin}, \citenamefont {Yang}, \citenamefont {Zheng}, \citenamefont {Shi},
  \citenamefont {Watanabe}, \citenamefont {Taniguchi}, \citenamefont {Tongay},
  \citenamefont {Zettl},\ and\ \citenamefont {Wang}}]{FengWang2019}%
  \BibitemOpen
  \bibfield  {author} {\bibinfo {author} {\bibfnamefont {C.}~\bibnamefont
  {Jin}}, \bibinfo {author} {\bibfnamefont {E.~C.}\ \bibnamefont {Regan}},
  \bibinfo {author} {\bibfnamefont {A.}~\bibnamefont {Yan}}, \bibinfo {author}
  {\bibfnamefont {M.}~\bibnamefont {Iqbal Bakti~Utama}}, \bibinfo {author}
  {\bibfnamefont {D.}~\bibnamefont {Wang}}, \bibinfo {author} {\bibfnamefont
  {S.}~\bibnamefont {Zhao}}, \bibinfo {author} {\bibfnamefont {Y.}~\bibnamefont
  {Qin}}, \bibinfo {author} {\bibfnamefont {S.}~\bibnamefont {Yang}}, \bibinfo
  {author} {\bibfnamefont {Z.}~\bibnamefont {Zheng}}, \bibinfo {author}
  {\bibfnamefont {S.}~\bibnamefont {Shi}}, \bibinfo {author} {\bibfnamefont
  {K.}~\bibnamefont {Watanabe}}, \bibinfo {author} {\bibfnamefont
  {T.}~\bibnamefont {Taniguchi}}, \bibinfo {author} {\bibfnamefont
  {S.}~\bibnamefont {Tongay}}, \bibinfo {author} {\bibfnamefont
  {A.}~\bibnamefont {Zettl}},\ and\ \bibinfo {author} {\bibfnamefont
  {F.}~\bibnamefont {Wang}},\ }\bibfield  {title} {\bibinfo {title}
  {Observation of moiré excitons in {WSe}$_2$/{WS}$_2$ heterostructure
  superlattices},\ }\href {https://doi.org/10.1038/s41586-019-0976-y}
  {\bibfield  {journal} {\bibinfo  {journal} {Nature}\ }\textbf {\bibinfo
  {volume} {567}},\ \bibinfo {pages} {76} (\bibinfo {year} {2019})}\BibitemShut
  {NoStop}%
\bibitem [{\citenamefont {Korm{\ifmmode\acute{a}\else\'{a}\fi}nyos}\ \emph
  {et~al.}(2015)\citenamefont {Korm{\ifmmode\acute{a}\else\'{a}\fi}nyos},
  \citenamefont {Burkard}, \citenamefont {Gmitra}, \citenamefont {Fabian},
  \citenamefont {Z{\ifmmode\acute{o}\else\'{o}\fi}lyomi}, \citenamefont
  {Drummond},\ and\ \citenamefont {Fal{'}ko}}]{Kormanyos2015}%
  \BibitemOpen
  \bibfield  {author} {\bibinfo {author} {\bibfnamefont {A.}~\bibnamefont
  {Korm{\ifmmode\acute{a}\else\'{a}\fi}nyos}}, \bibinfo {author} {\bibfnamefont
  {G.}~\bibnamefont {Burkard}}, \bibinfo {author} {\bibfnamefont
  {M.}~\bibnamefont {Gmitra}}, \bibinfo {author} {\bibfnamefont
  {J.}~\bibnamefont {Fabian}}, \bibinfo {author} {\bibfnamefont
  {V.}~\bibnamefont {Z{\ifmmode\acute{o}\else\'{o}\fi}lyomi}}, \bibinfo
  {author} {\bibfnamefont {N.~D.}\ \bibnamefont {Drummond}},\ and\ \bibinfo
  {author} {\bibfnamefont {V.}~\bibnamefont {Fal{'}ko}},\ }\bibfield  {title}
  {\bibinfo {title} {{k{$\cdot$}p theory for two-dimensional transition metal
  dichalcogenide semiconductors}},\ }\href
  {https://doi.org/10.1088/2053-1583/2/2/022001} {\bibfield  {journal}
  {\bibinfo  {journal} {2D Mater.}\ }\textbf {\bibinfo {volume} {2}},\ \bibinfo
  {pages} {022001} (\bibinfo {year} {2015})}\BibitemShut {NoStop}%
\bibitem [{\citenamefont {Goryca}\ \emph {et~al.}(2019)\citenamefont {Goryca},
  \citenamefont {Li}, \citenamefont {Stier}, \citenamefont {Taniguchi},
  \citenamefont {Watanabe}, \citenamefont {Courtade}, \citenamefont {Shree},
  \citenamefont {Robert}, \citenamefont {Urbaszek}, \citenamefont {Marie},\
  and\ \citenamefont {Crooker}}]{Goryca2019}%
  \BibitemOpen
  \bibfield  {author} {\bibinfo {author} {\bibfnamefont {M.}~\bibnamefont
  {Goryca}}, \bibinfo {author} {\bibfnamefont {J.}~\bibnamefont {Li}}, \bibinfo
  {author} {\bibfnamefont {A.~V.}\ \bibnamefont {Stier}}, \bibinfo {author}
  {\bibfnamefont {T.}~\bibnamefont {Taniguchi}}, \bibinfo {author}
  {\bibfnamefont {K.}~\bibnamefont {Watanabe}}, \bibinfo {author}
  {\bibfnamefont {E.}~\bibnamefont {Courtade}}, \bibinfo {author}
  {\bibfnamefont {S.}~\bibnamefont {Shree}}, \bibinfo {author} {\bibfnamefont
  {C.}~\bibnamefont {Robert}}, \bibinfo {author} {\bibfnamefont
  {B.}~\bibnamefont {Urbaszek}}, \bibinfo {author} {\bibfnamefont
  {X.}~\bibnamefont {Marie}},\ and\ \bibinfo {author} {\bibfnamefont {S.~A.}\
  \bibnamefont {Crooker}},\ }\bibfield  {title} {\bibinfo {title} {{Revealing
  exciton masses and dielectric properties of monolayer semiconductors with
  high magnetic fields}},\ }\href {https://doi.org/10.1038/s41467-019-12180-y}
  {\bibfield  {journal} {\bibinfo  {journal} {Nat. Commun.}\ }\textbf {\bibinfo
  {volume} {10}},\ \bibinfo {pages} {1} (\bibinfo {year} {2019})}\BibitemShut
  {NoStop}%
\bibitem [{\citenamefont {Wang}\ \emph
  {et~al.}(2018{\natexlab{b}})\citenamefont {Wang}, \citenamefont {Chiu},
  \citenamefont {Honz}, \citenamefont {Mak},\ and\ \citenamefont
  {Shan}}]{Mak_Shan_WSe2_IX_2018}%
  \BibitemOpen
  \bibfield  {author} {\bibinfo {author} {\bibfnamefont {Z.}~\bibnamefont
  {Wang}}, \bibinfo {author} {\bibfnamefont {Y.-H.}\ \bibnamefont {Chiu}},
  \bibinfo {author} {\bibfnamefont {K.}~\bibnamefont {Honz}}, \bibinfo {author}
  {\bibfnamefont {K.~F.}\ \bibnamefont {Mak}},\ and\ \bibinfo {author}
  {\bibfnamefont {J.}~\bibnamefont {Shan}},\ }\bibfield  {title} {\bibinfo
  {title} {{Electrical Tuning of Interlayer Exciton Gases in WSe2 Bilayers}},\
  }\href {https://doi.org/10.1021/acs.nanolett.7b03667} {\bibfield  {journal}
  {\bibinfo  {journal} {Nano Lett.}\ }\textbf {\bibinfo {volume} {18}},\
  \bibinfo {pages} {137} (\bibinfo {year} {2018}{\natexlab{b}})}\BibitemShut
  {NoStop}%
\bibitem [{\citenamefont {Al{\ifmmode\acute{e}\else\'{e}\fi}n}\ \emph
  {et~al.}(2003)\citenamefont {Al{\ifmmode\acute{e}\else\'{e}\fi}n},
  \citenamefont {Bickel}, \citenamefont {Karrai}, \citenamefont {Warburton},\
  and\ \citenamefont {Petroff}}]{modulation2003}%
  \BibitemOpen
  \bibfield  {author} {\bibinfo {author} {\bibfnamefont {B.}~\bibnamefont
  {Al{\ifmmode\acute{e}\else\'{e}\fi}n}}, \bibinfo {author} {\bibfnamefont
  {F.}~\bibnamefont {Bickel}}, \bibinfo {author} {\bibfnamefont
  {K.}~\bibnamefont {Karrai}}, \bibinfo {author} {\bibfnamefont {R.~J.}\
  \bibnamefont {Warburton}},\ and\ \bibinfo {author} {\bibfnamefont {P.~M.}\
  \bibnamefont {Petroff}},\ }\bibfield  {title} {\bibinfo {title} {{Stark-shift
  modulation absorption spectroscopy of single quantum dots}},\ }\href
  {https://doi.org/10.1063/1.1609243} {\bibfield  {journal} {\bibinfo
  {journal} {Appl. Phys. Lett.}\ }\textbf {\bibinfo {volume} {83}},\ \bibinfo
  {pages} {2235} (\bibinfo {year} {2003})}\BibitemShut {NoStop}%
\bibitem [{\citenamefont {H{\ifmmode\ddot{o}\else\"{o}\fi}gele}\ \emph
  {et~al.}(2004)\citenamefont {H{\ifmmode\ddot{o}\else\"{o}\fi}gele},
  \citenamefont {Al{\ifmmode\grave{e}\else\`{e}\fi}n}, \citenamefont {Bickel},
  \citenamefont {Warburton}, \citenamefont {Petroff},\ and\ \citenamefont
  {Karrai}}]{Hoegele2004}%
  \BibitemOpen
  \bibfield  {author} {\bibinfo {author} {\bibfnamefont {A.}~\bibnamefont
  {H{\ifmmode\ddot{o}\else\"{o}\fi}gele}}, \bibinfo {author} {\bibfnamefont
  {B.}~\bibnamefont {Al{\ifmmode\grave{e}\else\`{e}\fi}n}}, \bibinfo {author}
  {\bibfnamefont {F.}~\bibnamefont {Bickel}}, \bibinfo {author} {\bibfnamefont
  {R.~J.}\ \bibnamefont {Warburton}}, \bibinfo {author} {\bibfnamefont {P.~M.}\
  \bibnamefont {Petroff}},\ and\ \bibinfo {author} {\bibfnamefont
  {K.}~\bibnamefont {Karrai}},\ }\bibfield  {title} {\bibinfo {title} {{Exciton
  fine structure splitting of single InGaAs self-assembled quantum dots}},\
  }\href {https://doi.org/10.1016/j.physe.2003.11.007} {\bibfield  {journal}
  {\bibinfo  {journal} {Physica E}\ }\textbf {\bibinfo {volume} {21}},\
  \bibinfo {pages} {175} (\bibinfo {year} {2004})}\BibitemShut {NoStop}%
\bibitem [{\citenamefont {Barré}\ \emph {et~al.}(2022)\citenamefont {Barré},
  \citenamefont {Karni}, \citenamefont {Liu}, \citenamefont {O’Beirne},
  \citenamefont {Chen}, \citenamefont {Ribeiro}, \citenamefont {Yu},
  \citenamefont {Kim}, \citenamefont {Watanabe}, \citenamefont {Taniguchi},
  \citenamefont {Barmak}, \citenamefont {Lui}, \citenamefont
  {Refaely-Abramson}, \citenamefont {da~Jornada},\ and\ \citenamefont
  {Heinz}}]{BarreHeinz2022}%
  \BibitemOpen
  \bibfield  {author} {\bibinfo {author} {\bibfnamefont {E.}~\bibnamefont
  {Barré}}, \bibinfo {author} {\bibfnamefont {O.}~\bibnamefont {Karni}},
  \bibinfo {author} {\bibfnamefont {E.}~\bibnamefont {Liu}}, \bibinfo {author}
  {\bibfnamefont {A.~L.}\ \bibnamefont {O’Beirne}}, \bibinfo {author}
  {\bibfnamefont {X.}~\bibnamefont {Chen}}, \bibinfo {author} {\bibfnamefont
  {H.~B.}\ \bibnamefont {Ribeiro}}, \bibinfo {author} {\bibfnamefont
  {L.}~\bibnamefont {Yu}}, \bibinfo {author} {\bibfnamefont {B.}~\bibnamefont
  {Kim}}, \bibinfo {author} {\bibfnamefont {K.}~\bibnamefont {Watanabe}},
  \bibinfo {author} {\bibfnamefont {T.}~\bibnamefont {Taniguchi}}, \bibinfo
  {author} {\bibfnamefont {K.}~\bibnamefont {Barmak}}, \bibinfo {author}
  {\bibfnamefont {C.~H.}\ \bibnamefont {Lui}}, \bibinfo {author} {\bibfnamefont
  {S.}~\bibnamefont {Refaely-Abramson}}, \bibinfo {author} {\bibfnamefont
  {F.~H.}\ \bibnamefont {da~Jornada}},\ and\ \bibinfo {author} {\bibfnamefont
  {T.~F.}\ \bibnamefont {Heinz}},\ }\bibfield  {title} {\bibinfo {title}
  {Optical absorption of interlayer excitons in transition-metal dichalcogenide
  heterostructures},\ }\href {https://doi.org/10.1126/science.abm8511}
  {\bibfield  {journal} {\bibinfo  {journal} {Science}\ }\textbf {\bibinfo
  {volume} {376}},\ \bibinfo {pages} {406} (\bibinfo {year}
  {2022})}\BibitemShut {NoStop}%
\bibitem [{\citenamefont {Faria~Junior}\ and\ \citenamefont
  {Fabian}(2023)}]{Paulo_2023}%
  \BibitemOpen
  \bibfield  {author} {\bibinfo {author} {\bibfnamefont {P.~E.}\ \bibnamefont
  {Faria~Junior}}\ and\ \bibinfo {author} {\bibfnamefont {J.}~\bibnamefont
  {Fabian}},\ }\bibfield  {title} {\bibinfo {title} {Signatures of electric
  field and layer separation effects on the spin-valley physics of mose2/wse2
  heterobilayers: From energy bands to dipolar excitons},\ }\bibfield
  {journal} {\bibinfo  {journal} {Nanomaterials}\ }\textbf {\bibinfo {volume}
  {13}},\ \href {https://doi.org/10.3390/nano13071187} {10.3390/nano13071187}
  (\bibinfo {year} {2023})\BibitemShut {NoStop}%
\bibitem [{\citenamefont {Gobato}\ \emph {et~al.}(2022)\citenamefont {Gobato},
  \citenamefont {de~Brito}, \citenamefont {Chaves}, \citenamefont {Prosnikov},
  \citenamefont {Wo{\ifmmode\acute{z}\else\'{z}\fi}niak}, \citenamefont {Guo},
  \citenamefont {Barcelos}, \citenamefont
  {Milo{\ifmmode\check{s}\else\v{s}\fi}evi{\ifmmode\acute{c}\else\'{c}\fi}},
  \citenamefont {Withers},\ and\ \citenamefont
  {Christianen}}]{Gobato_Gfactors2022}%
  \BibitemOpen
  \bibfield  {author} {\bibinfo {author} {\bibfnamefont {Y.~G.}\ \bibnamefont
  {Gobato}}, \bibinfo {author} {\bibfnamefont {C.~S.}\ \bibnamefont
  {de~Brito}}, \bibinfo {author} {\bibfnamefont {A.}~\bibnamefont {Chaves}},
  \bibinfo {author} {\bibfnamefont {M.~A.}\ \bibnamefont {Prosnikov}}, \bibinfo
  {author} {\bibfnamefont {T.}~\bibnamefont
  {Wo{\ifmmode\acute{z}\else\'{z}\fi}niak}}, \bibinfo {author} {\bibfnamefont
  {S.}~\bibnamefont {Guo}}, \bibinfo {author} {\bibfnamefont {I.~D.}\
  \bibnamefont {Barcelos}}, \bibinfo {author} {\bibfnamefont {M.~V.}\
  \bibnamefont
  {Milo{\ifmmode\check{s}\else\v{s}\fi}evi{\ifmmode\acute{c}\else\'{c}\fi}}},
  \bibinfo {author} {\bibfnamefont {F.}~\bibnamefont {Withers}},\ and\ \bibinfo
  {author} {\bibfnamefont {P.~C.~M.}\ \bibnamefont {Christianen}},\ }\bibfield
  {title} {\bibinfo {title} {{Distinctive g-Factor of
  Moir{\ifmmode\acute{e}\else\'{e}\fi}-Confined Excitons in van der Waals
  Heterostructures}},\ }\href {https://doi.org/10.1021/acs.nanolett.2c03008}
  {\bibfield  {journal} {\bibinfo  {journal} {Nano Lett.}\ }\textbf {\bibinfo
  {volume} {22}},\ \bibinfo {pages} {8641} (\bibinfo {year}
  {2022})}\BibitemShut {NoStop}%
\bibitem [{\citenamefont {Naik}\ \emph {et~al.}(2022)\citenamefont {Naik},
  \citenamefont {Regan}, \citenamefont {Zhang}, \citenamefont {Chan},
  \citenamefont {Li}, \citenamefont {Wang}, \citenamefont {Yoon}, \citenamefont
  {Ong}, \citenamefont {Zhao}, \citenamefont {Zhao}, \citenamefont {Utama},
  \citenamefont {Gao}, \citenamefont {Wei}, \citenamefont {Sayyad},
  \citenamefont {Yumigeta}, \citenamefont {Watanabe}, \citenamefont
  {Taniguchi}, \citenamefont {Tongay}, \citenamefont {da~Jornada},
  \citenamefont {Wang},\ and\ \citenamefont {Louie}}]{naik_intralayer_2022}%
  \BibitemOpen
  \bibfield  {author} {\bibinfo {author} {\bibfnamefont {M.~H.}\ \bibnamefont
  {Naik}}, \bibinfo {author} {\bibfnamefont {E.~C.}\ \bibnamefont {Regan}},
  \bibinfo {author} {\bibfnamefont {Z.}~\bibnamefont {Zhang}}, \bibinfo
  {author} {\bibfnamefont {Y.-H.}\ \bibnamefont {Chan}}, \bibinfo {author}
  {\bibfnamefont {Z.}~\bibnamefont {Li}}, \bibinfo {author} {\bibfnamefont
  {D.}~\bibnamefont {Wang}}, \bibinfo {author} {\bibfnamefont {Y.}~\bibnamefont
  {Yoon}}, \bibinfo {author} {\bibfnamefont {C.~S.}\ \bibnamefont {Ong}},
  \bibinfo {author} {\bibfnamefont {W.}~\bibnamefont {Zhao}}, \bibinfo {author}
  {\bibfnamefont {S.}~\bibnamefont {Zhao}}, \bibinfo {author} {\bibfnamefont
  {M.~I.~B.}\ \bibnamefont {Utama}}, \bibinfo {author} {\bibfnamefont
  {B.}~\bibnamefont {Gao}}, \bibinfo {author} {\bibfnamefont {X.}~\bibnamefont
  {Wei}}, \bibinfo {author} {\bibfnamefont {M.}~\bibnamefont {Sayyad}},
  \bibinfo {author} {\bibfnamefont {K.}~\bibnamefont {Yumigeta}}, \bibinfo
  {author} {\bibfnamefont {K.}~\bibnamefont {Watanabe}}, \bibinfo {author}
  {\bibfnamefont {T.}~\bibnamefont {Taniguchi}}, \bibinfo {author}
  {\bibfnamefont {S.}~\bibnamefont {Tongay}}, \bibinfo {author} {\bibfnamefont
  {F.~H.}\ \bibnamefont {da~Jornada}}, \bibinfo {author} {\bibfnamefont
  {F.}~\bibnamefont {Wang}},\ and\ \bibinfo {author} {\bibfnamefont {S.~G.}\
  \bibnamefont {Louie}},\ }\bibfield  {title} {\bibinfo {title} {Intralayer
  charge-transfer moiré excitons in van der {Waals} superlattices},\ }\href
  {https://doi.org/10.1038/s41586-022-04991-9} {\bibfield  {journal} {\bibinfo
  {journal} {Nature}\ }\textbf {\bibinfo {volume} {609}},\ \bibinfo {pages}
  {52} (\bibinfo {year} {2022})}\BibitemShut {NoStop}%
\bibitem [{\citenamefont {Arora}\ \emph {et~al.}(2013)\citenamefont {Arora},
  \citenamefont {Mandal}, \citenamefont {Chakrabarti},\ and\ \citenamefont
  {Ghosh}}]{arora2013magneto}%
  \BibitemOpen
  \bibfield  {author} {\bibinfo {author} {\bibfnamefont {A.}~\bibnamefont
  {Arora}}, \bibinfo {author} {\bibfnamefont {A.}~\bibnamefont {Mandal}},
  \bibinfo {author} {\bibfnamefont {S.}~\bibnamefont {Chakrabarti}},\ and\
  \bibinfo {author} {\bibfnamefont {S.}~\bibnamefont {Ghosh}},\ }\bibfield
  {title} {\bibinfo {title} {{Magneto-optical Kerr effect spectroscopy based
  study of Land{\ifmmode\acute{e}\else\'{e}\fi} g-factor for holes in
  GaAs/AlGaAs single quantum wells under low magnetic fields}},\ }\href
  {https://doi.org/10.1063/1.4808302} {\bibfield  {journal} {\bibinfo
  {journal} {J. Appl. Phys.}\ }\textbf {\bibinfo {volume} {113}},\ \bibinfo
  {pages} {213505} (\bibinfo {year} {2013})}\BibitemShut {NoStop}%
\bibitem [{\citenamefont {Zhao}\ \emph {et~al.}(2023)\citenamefont {Zhao},
  \citenamefont {Li}, \citenamefont {Huang}, \citenamefont {Rupp},
  \citenamefont {G{\ifmmode\ddot{o}\else\"{o}\fi}ser}, \citenamefont {Vovk},
  \citenamefont {Kruchinin}, \citenamefont {Watanabe}, \citenamefont
  {Taniguchi}, \citenamefont {Bilgin}, \citenamefont {Baimuratov},\ and\
  \citenamefont {H{\ifmmode\ddot{o}\else\"{o}\fi}gele}}]{Zhao2023}%
  \BibitemOpen
  \bibfield  {author} {\bibinfo {author} {\bibfnamefont {S.}~\bibnamefont
  {Zhao}}, \bibinfo {author} {\bibfnamefont {Z.}~\bibnamefont {Li}}, \bibinfo
  {author} {\bibfnamefont {X.}~\bibnamefont {Huang}}, \bibinfo {author}
  {\bibfnamefont {A.}~\bibnamefont {Rupp}}, \bibinfo {author} {\bibfnamefont
  {J.}~\bibnamefont {G{\ifmmode\ddot{o}\else\"{o}\fi}ser}}, \bibinfo {author}
  {\bibfnamefont {I.~A.}\ \bibnamefont {Vovk}}, \bibinfo {author}
  {\bibfnamefont {S.~{\relax Yu}.}\ \bibnamefont {Kruchinin}}, \bibinfo
  {author} {\bibfnamefont {K.}~\bibnamefont {Watanabe}}, \bibinfo {author}
  {\bibfnamefont {T.}~\bibnamefont {Taniguchi}}, \bibinfo {author}
  {\bibfnamefont {I.}~\bibnamefont {Bilgin}}, \bibinfo {author} {\bibfnamefont
  {A.~S.}\ \bibnamefont {Baimuratov}},\ and\ \bibinfo {author} {\bibfnamefont
  {A.}~\bibnamefont {H{\ifmmode\ddot{o}\else\"{o}\fi}gele}},\ }\bibfield
  {title} {\bibinfo {title} {{Excitons in mesoscopically reconstructed
  moir{\ifmmode\acute{e}\else\'{e}\fi} heterostructures}},\ }\href
  {https://doi.org/10.1038/s41565-023-01356-9} {\bibfield  {journal} {\bibinfo
  {journal} {Nat. Nanotechnol.}\ }\textbf {\bibinfo {volume} {18}},\ \bibinfo
  {pages} {572} (\bibinfo {year} {2023})}\BibitemShut {NoStop}%
\bibitem [{\citenamefont {McIntyre}\ and\ \citenamefont
  {Aspnes}(1971)}]{mcintyre1971differential}%
  \BibitemOpen
  \bibfield  {author} {\bibinfo {author} {\bibfnamefont {J.~D.~E.}\
  \bibnamefont {McIntyre}}\ and\ \bibinfo {author} {\bibfnamefont {D.~E.}\
  \bibnamefont {Aspnes}},\ }\bibfield  {title} {\bibinfo {title} {{Differential
  reflection spectroscopy of very thin surface films}},\ }\href
  {https://doi.org/10.1016/0039-6028(71)90272-X} {\bibfield  {journal}
  {\bibinfo  {journal} {Surf. Sci.}\ }\textbf {\bibinfo {volume} {24}},\
  \bibinfo {pages} {417} (\bibinfo {year} {1971})}\BibitemShut {NoStop}%

  
  
\end{thebibliography}

%

\end{document}


\title{Supplementary Material: \\ Field-induced hybridization of moir\'e excitons in MoSe$_2$/WS$_2$ heterobilayers}

\author{Borislav Polovnikov}
\author{Johannes Scherzer}
\author{Subhradeep Misra}
\author{Xin Huang}
\author{Christian Mohl}
\author{Zhijie Li}
\author{Jonas G{\"o}ser}
\author{Jonathan F\"orste}
\author{Ismail Bilgin}
\author{Kenji Watanabe}
\author{Takashi Taniguchi}
\author{Alexander H{\"o}gele}
\author{Anvar~S.~Baimuratov}

\maketitle
\vspace{15pt}

\section{Device layout and fabrication}
\begin{figure}[h]  
\includegraphics[width=0.5\columnwidth]{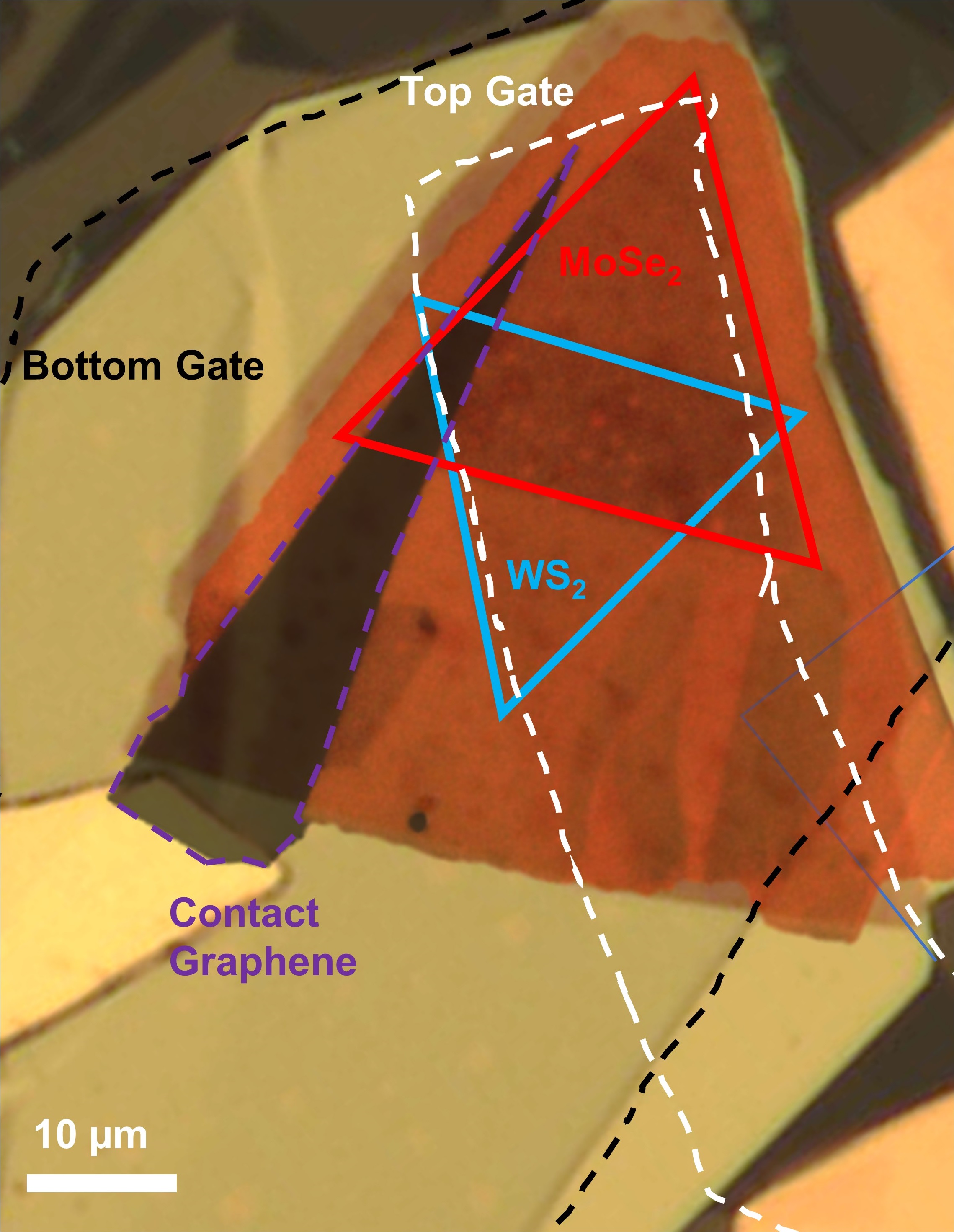}
\caption{Optical image of the studied MoSe$_2$/WS$_2$ device. The triangular single-crystal monolayers of MoSe$_2$ and WS$_2$ have a twist angle of approximately $179^\circ$ and are delimited in red and blue. The few-layer graphite flakes used for the electric gates are delimited by dashed lines.}
\label{device}
\end{figure}

We obtained monolayers of MoSe$_2$ and WS$_2$ from in-house CVD synthesis. Thin flakes of hBN were exfoliated from bulk crystals provided by NIMS, Japan. All flakes were then stacked together by the dry-transfer technique using PDMS/PC stamps, deposited onto a Si/SiO$_2$ substrate and annealed at 200°C for 12 hours. Given the triangular shape of CVD-grown TMD crystals we controlled the relative orientation of the two crystals and fabricated an H-type sample (anti parallel alignment) with a rotation angle of ca. $179^\circ$, see Fig.~\ref{device}.

%
%

\newpage
\section{Calculation of the absorption coefficient $\chi''$}
\begin{figure}[h!]  
\begin{center}
\includegraphics[scale=0.8]{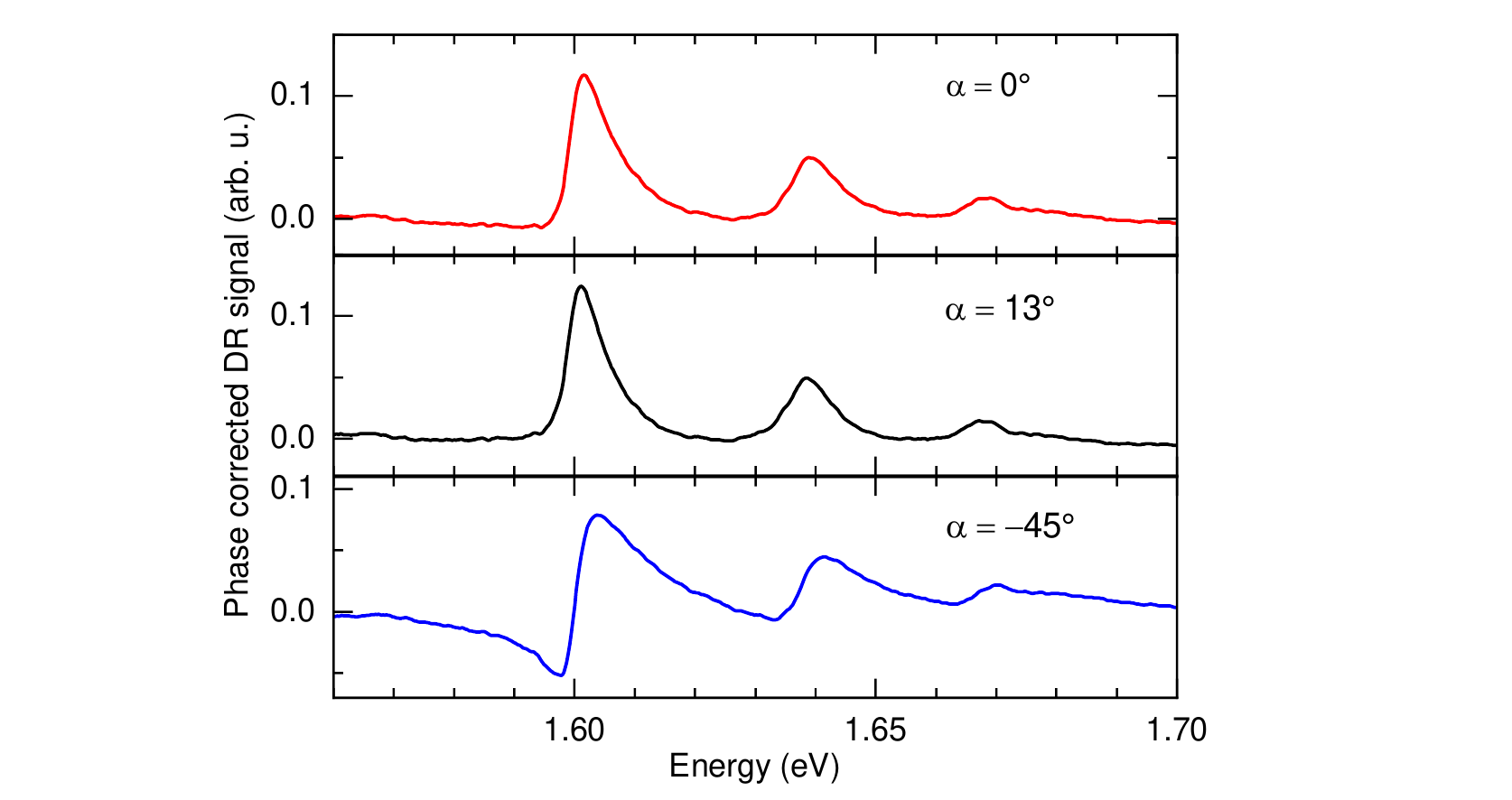}
\caption{Top: Raw DR signal corresponding to the data shown in Fig. 1b of the main text. Middle: Phase corrected DR with the phase $\alpha = 13^\circ$ which was used in the main text. Bottom: Phase corrected DR with $\alpha = -45^\circ$. }
\label{KramersKronig}
\end{center}
\end{figure}

For comparison between measurements and theory we translate the differential reflectance (DR) signal into the imaginary part of the dielectric susceptibility $\chi = \chi' + i\chi''$ by following Refs.~\onlinecite{arora2013magneto,Zhao2023}. We define DR as $(R_0 - R)/R_0$, where $R$ is the reflectance signal from the sample and $R_0$ is a reference background signal from the substrate. As pointed out in~\cite{mcintyre1971differential}, without hBN encapsulation the DR signal would simply be proportional to $\chi''$. In the present device, however, the top and bottom hBN (both of $55$~nm thickness) add additional interfaces that influence the lineshape of the reflected signal by interference effects. As shown in~\cite{arora2013magneto}, these effects can be taken into account by introducing a phenomenological phase factor for the effective susceptibility $\tilde\chi = \chi \, e^{i\alpha}$, where $\alpha$ is an \textit{à priori} wavelength-dependent phase factor and the DR signal is now proportional to the imaginary part of $\tilde\chi$, $\text{DR} \sim \text{Im}(\tilde\chi)$. Here, for simplicity, we assume $\alpha$ to be a constant in the relevant spectral range. Then, using the Kramers-Kronig relations for the complex-valued function $\tilde\chi (\omega)$, we can compute $\chi''$ from DR as
\begin{equation}
\chi''(\omega) = \text{Im}\left(e^{-i\alpha} \tilde\chi (\omega)\right) = \cos{(\alpha)} \, \text{DR}(\omega) - \sin{(\alpha)} \, \int_\mathbb{R} \,  \frac{\text{DR}(\omega')}{\omega - \omega'} \, d\omega' \, .
\end{equation}
In Fig.~\ref{KramersKronig} we plot the raw DR signal corresponding to the data shown in Fig. 1b of the main text ($\alpha = 0^\circ$) alongside the phase-corrected data ($\alpha = 13^\circ$) and a third example with an arbitrary phase for illustration of the principle ($\alpha = -45^\circ$). The phase which was used in the main text is chosen such that the individual moir\'e exciton peaks exhibit almost lorentzian lineshape, as expected for the absorption of individual exciton resonances.
%
%

\section{Experimental Setup}
Cryogenic spectroscopy was conducted using home-built confocal microscopes in back-scattering geometry. The sample was loaded into a closed-cycle cryostat (attocube systems, attoDRY1000) with a base temperature of $3.2$~K equipped with a superconducting magnet providing magnetic fields of up to $\pm 9$~T in Faraday configuration. Piezo-stepping and scanning units (attocube systems, ANPxyz and ANSxy100) were used for sample positioning with respect to a low-temperature apochromatic objective (attocube systems). 

For PL and white light DR, the signal was spectrally dispersed by a monochromator (Acton SP2500 with a 300 grooves/mm grating) and detected by a liquid nitrogen cooled charge-coupled device (Spec-10:100BR). A set of linear polarizers (Thorlabs, LPVIS), half- and quarter-waveplates (B. Halle, $310-1100$~nm achromatic) mounted on piezo-rotators (attocube systems, ANR240) were used to control the polarization in excitation and detection. A wavelength-tunable Ti:sapphire laser (Mira) in continuous-wave mode and laser diodes were used to excite PL. For DR measurements, a stabilized Tungsten-Halogen lamp (Thorlabs, SLS201L) and supercontinuum lasers (NKT Photonics, SuperK Extreme and SuperK Varia) were used as broadband light sources.

\begin{figure}[h!]  
\begin{center}
\includegraphics[width=0.7\columnwidth]{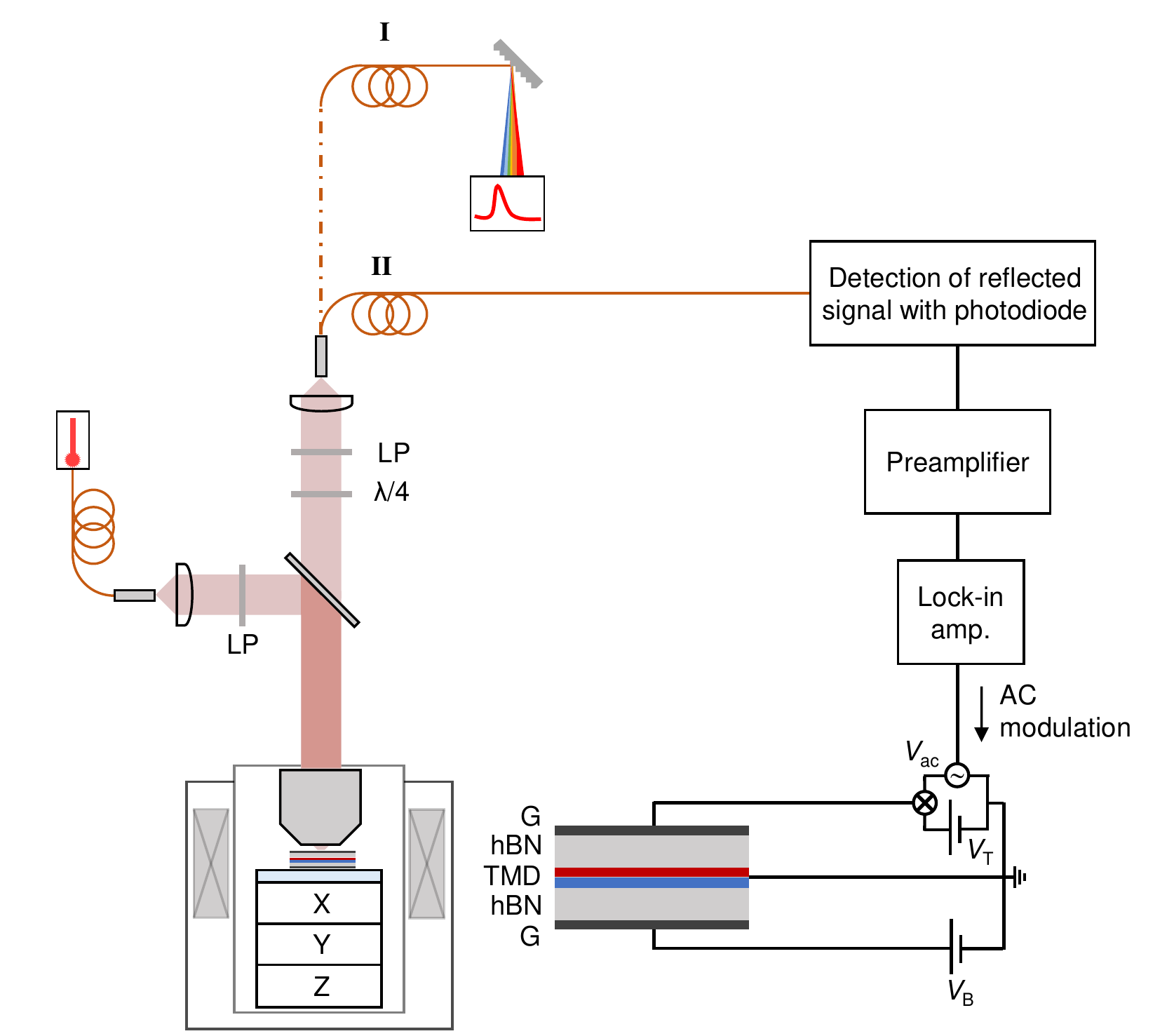}
\caption{Experimental setup for PL and DR measurements~(I) and for narrow band modulation spectroscopy~(II). The sample is mounted on a stack of piezo positioners (X/Y/Z) inside a closed-cycle cryostat at a base temperature of $3.2$~K. A home built, fiber-based confocal microscope with an apochromatic objective is used to focus the excitation light (horizontal microscope arm) onto the sample, and the reflected/emitted light is then collected by the same objective and coupled into the detection fiber (vertical microscope arm). The light from the detection fiber is either dispersed by a monochromator for spectral information (I), or detected by a silicon photodiode for narrow-band modulation measurements (II). A set of  linear polarizers (LP), waveplates ($\lambda / 4$) and optionally also optical filters is used to  control the incident light.}
\label{Setup}
\end{center}
\end{figure}

For narrow band modulation spectroscopy, a wavelength-tunable Ti:sapphire laser (M squared) of $50$~$\mu$eV linewidth was used to excite the sample while modulating the top gate with an ac-voltage $V_{\text{ac}}=500$~mV at $147$~Hz. The reflected optical signal was detected by a silicon photodiode and its output was preamplified by Ithaco Model 1211. The amplified dc part of the photosignal ($R_{\text{dc}}$) was measured, and the ac part was demodulated and amplified further by lock-in amplifier (EG\&G 7260), supplied with the reference frequency same as that of the modulation voltage to obtain $R_{\text{ac}}$ (see Fig.~\ref{Setup}). Finally, the \textit{narrow-band differential reflectance} (DR$^\prime$) is given by the ratio of the demodulated ac part to the dc photosignal, $R_{\text{ac}}$/$R_{\text{dc}}$. An integration time constant of $1$~s was used per data point acquisition. In simple limits of the modulation conditions, the signal is proportional to the derivative of DR~\cite{modulation2003}, whereas in general, it requires a microscopic analysis including optical pumping of charge carrier reservoirs.

%

\newpage
\section{Measurement of exciton \lowercase{g}-factors}
\begin{figure}[h!]  
\begin{center}
\includegraphics[scale=1.0]{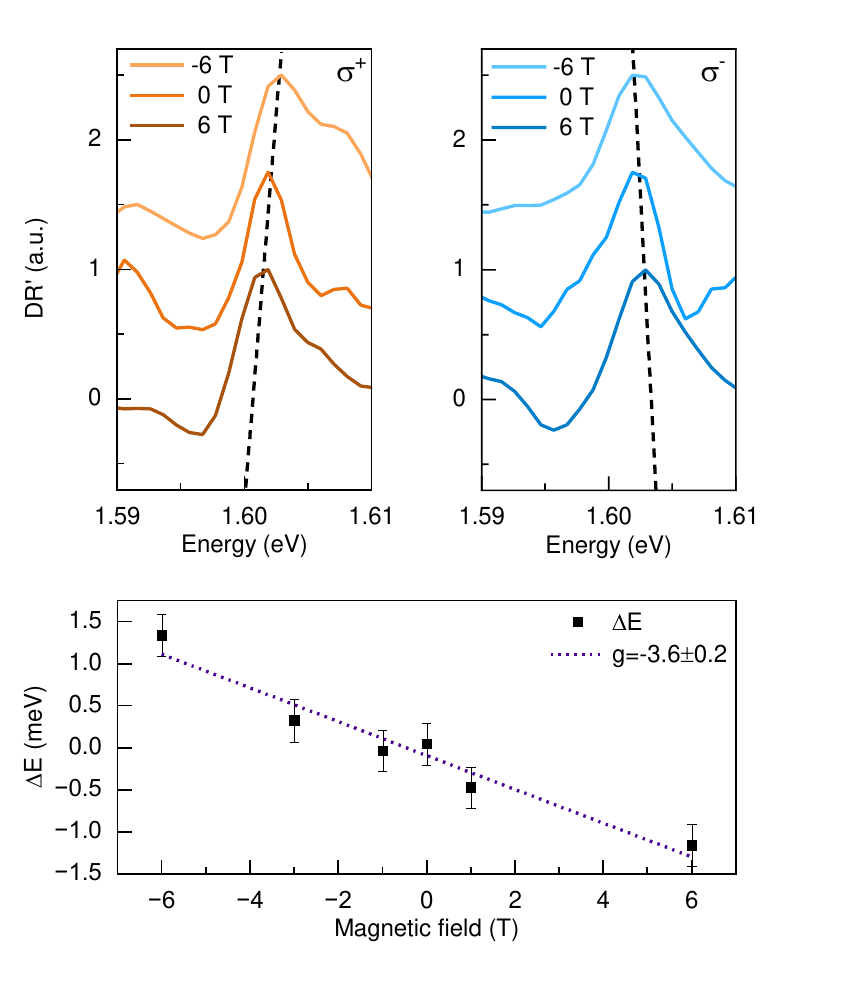}
\caption{Measurement procedure of the exciton g-factors. Top panel: representative spectra of the narrow-band modulation signal as obtained in circular polarization ($\sigma^+$ and $\sigma^-$) for linearly polarized excitation at magnetic fields between $B=-6$~T and $B=6$~T. For each data point, the peak energy was obtained from the maximum of the signal, where an uncertainty of $10\%$ of the linewidth was assumed. Bottom panel: the valley Zeeman splitting between the peak energies in the two polarization branches determines $\Delta E$. The g-factor was obtained from the linear slope of the fit of $\Delta E$ as a function of $B$.}
\label{g-factors}
\end{center}
\end{figure}

To measure the g-factors shown in Fig.~3b of the main text, we repeated the narrow-band modulation spectroscopy measurement for a series of out-of-plane magnetic fields between $B=6$~T and $B=-6$~T. The measurements were performed under linearly polarized excitation with detection in both $\sigma^+$ and $\sigma^-$ polarization. The valley Zeeman energy splitting $\Delta E$ between the two polarization branches was obtained from the peak-maxima energies, and the g-factor was extracted from the slope of the best linear fit of $\Delta E$ as a function of $B$ as shown in Fig.~\ref{g-factors}.

%

To calculate the g-factors of moir\'e excitons one needs to determine the g-factor dependence on the wave vector in the corresponding monolayers. For large momentum mismatch $\Delta \mathbf{K}$, this dependence cannot be simply approximated by a paraboloid at the $\mathbf{K}$ or $\mathbf{K'}$ points. Thus, the g-factor of the moir\'e exciton \textit{m} is written as
\begin{equation}
g_m = \sum_l \sum_\mathbf{K} \sum_\mathbf{k}c_{l,m}(\mathbf{K}) \, \phi_\mathbf{k} \, \left[ g_{l, e}(\mathbf{k}, \mathbf{K})+g_{l,h}(\mathbf{k},\mathbf{K})\right] ,
\end{equation}
where we sum over the two layers $l=1,2$ and the vectors of COM and relative motion $\mathbf{K}$ and $\mathbf{k}$. Here, $\phi_\mathbf{k}$ is the relative wavefunction, $c_{l,m}(\mathbf{K})$ are coefficients related to the plane-wave projections $\left< j(l, \mathbf{K}) | m\right>$, and $g_{l, e}(\mathbf{k}, \mathbf{K})$ and $g_{l, h}(\mathbf{k}, \mathbf{K})$ are g-factors of the electron and hole.
\newpage
\section{Real space distribution of moir\'e exciton states}
To compute the real-space distribution of the emerging moir\'e excitons we denote the original basis of the Hamiltonian (1) in the main text by $|j\rangle$ with $j=0,...,12$, s.t. it holds e.g. $\langle j | H |j \rangle \, = \, E_j$ for $j=0,...,6$ and  $\langle j | H |j \rangle \, = \, \mathcal{E}_{j-7}$ for $j=7,...,12$. Similarly, the Hamiltonian's eigenbasis corresponding to the moir\'e excitons is denoted by $|m\rangle$.

We define $\mathbf{b}_0 = \mathbf{0}$ such that the intralayer states $|j\rangle$, $j=0,...,6$ correspond to the plane waves $\exp{(i\mathbf{b}_j \mathbf{r})}$. Then, the intralayer distribution of the moir\'e excitons $|m\rangle$ is simply computed by the plane-wave projections of these first seven states, \textit{i.e.}, 
$\psi_m(\mathbf{r})  = \sum_{j=0}^{6} \langle j | m \rangle \, e^{i \mathbf{b}_j \mathbf{r}}$. Using this formula for the bright moir\'e excitons results in the plots in Fig.~4 of the main text.

\begin{figure}[h!]  
\begin{center}
\includegraphics[width=1\columnwidth]{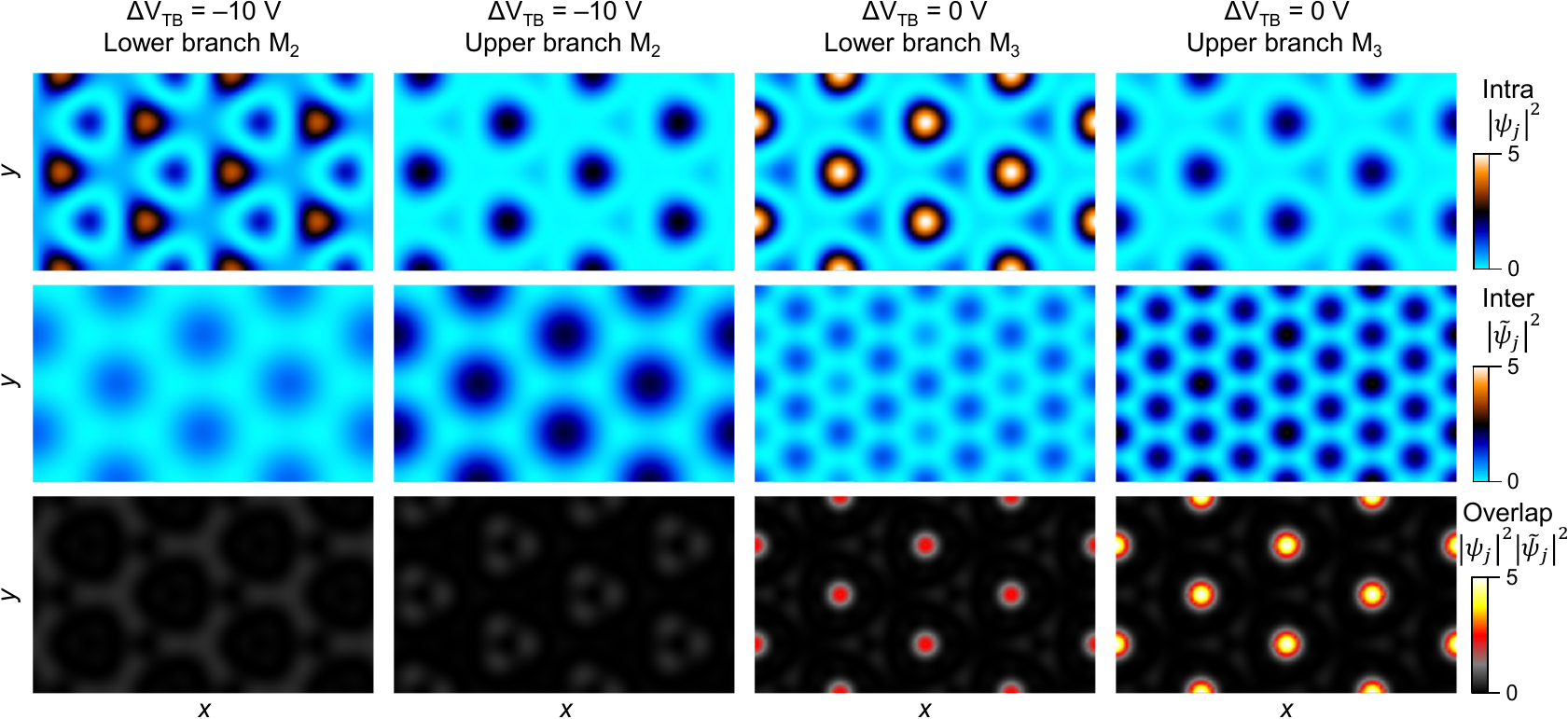}
\caption{The upper and middle panels show spatial distributions of intra- and interlayer exciton wave functions of the peaks $M_2(-10\text{V})$ and $M_3$, respectively. The bottom panel shows their wave functions overlap with the strongest overlap for the $M_3$ doublet.}
\label{interlayer}
\end{center}
\end{figure}

Similarly, the interlayer distribution is obtained from projections on the six interlayer states, $\tilde{\psi}_m(\mathbf{r})  = \sum_{j=7}^{12} \langle j | m \rangle \, e^{i \mathbf{b}_j \mathbf{r}}$, where we define $\mathbf{b}_j$ as the vectors pointing to the blue states defined in Fig. 1 of the manuscript (cp. Eq. 2 of the main text). Interestingly, comparing the intra- and interlayer distributions helps to provide an intuition on the coupling strengths between intra- and interlayer excitons which we specify for the peaks $M_2(-10\text{V})$ and $M_3$ in Fig.~\ref{interlayer}. If we plot the overlap of the respective wave functions for each branch as shown in the bottom panel, one easily sees that the overlap for the $M_3$ doublet is much larger than for the $M_2$ peak, resulting in the strongest anti-crossing as observed in Fig.~3 of the main text.\\
%
\newpage
\section{Limiting cases of the effective model}
The phenomenological model developed in the main text presents a combination of intralayer exciton mixing mediated by the potential $V$ and interlayer hybridization mediated by the hopping parameter $t$. Conversely, setting $V=0$~meV recovers the pure interlayer hybridization model, and setting $t=0$~meV results in a pure intralayer exciton mixing description.

\begin{figure}[h!]  
\begin{center}
\includegraphics[width=0.9\columnwidth]{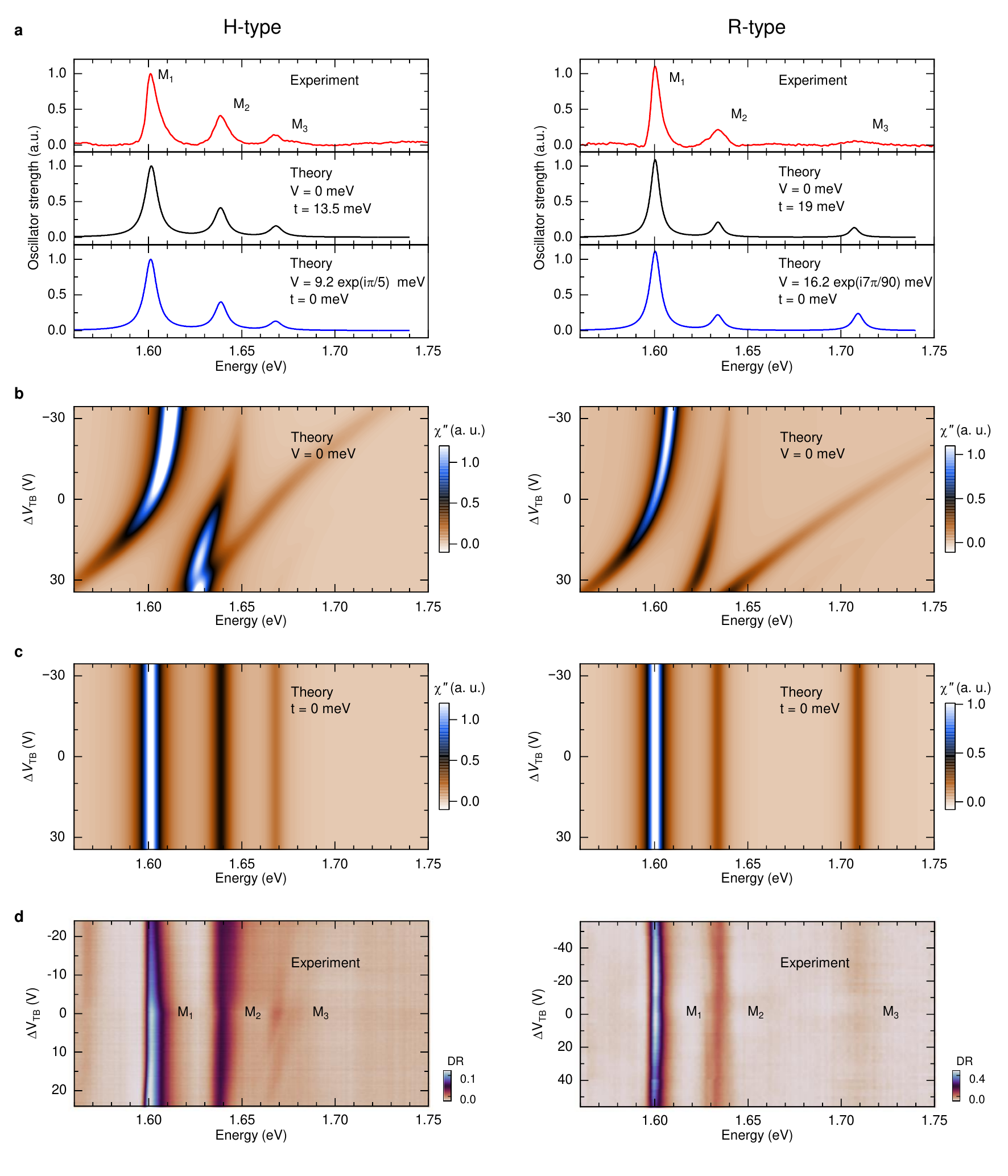}
\caption{Comparison of experimental data and modeling by the limiting cases $V = 0$~meV  and $t = 0$~meV, only. The left column refers to H-type data as shown in the main text, and the right column to R-type data. \textbf{a}. Experimental DR spectra with the three bright moiré excitons (note that in the R-stacking, $M_3$ appears $45$~meV blue-shifted with respect to $M_3$ in the H-stacking, which is a consequence of the reversed ordering of the spin-polarized conduction sub-bands), complemented with theoretical fits of the neutral spectra within the interlayer hybridization model ($V = 0$~meV) or within the intralayer mixing model ($t = 0$~meV) only. For the neutral regime, very good agreement between experiment and both theories can be obtained. \textbf{b} \& \textbf{c}. Calculated dispersions under perpendicularly applied electric fields for $V = 0$~meV (\textbf{b}) and $t = 0$~meV (\textbf{c}). \textbf{d}. Actual experimental dispersions of the three bright moiré excitons. In particular, there is no sizable energetic shifts for $M_1$ and $M_2$ in both stacking configurations, contrary to what the hybridization model predicts. However, $M_3$ presents an avoided crossing with a sizable Stark shift in the H-type data, contrary to the $t = 0$~meV simulation.}
\label{models}
\end{center}
\end{figure}

To assess these two descriptions, we fabricated a second sample in R-type stacking configuration. In Fig.~\ref{models}a, we show the experimental DR signals of both samples together with theoretical fits in both limiting cases of the model. The good agreement between theory and experiment shows that both the $V=0$~meV and $t=0$~meV subspaces provide enough descriptive power to model the neutral spectra of both H- and R-type devices. Therefore, further experimental data is needed in order to determine the correct contributions to the intra- and interlayer moir\'e potentials.

Notably, in the $V=0$~meV description, any exciton mixing is mediated through second-order hopping processes and is associated with  substantial out-of-plane electric dipoles as shown in Fig.~\ref{models}b. To describe the vanishing electric-field dispersion of the bright excitons as observed in experiment, it is therefore necessary to include the potential $V$ to enable first-order intralayer coupling. The strong avoided crossing of $M_3$ in the H-type data, on the contrary, requires finite interlayer hybridization. Therefore, to correctly describe all three bright excitons, it is necessary to combine both intralayer coupling and interlayer hybridization as shown in Fig.~2b of the main text.

\begin{figure}[h!]  
\begin{center}
\includegraphics[width=0.9\columnwidth]{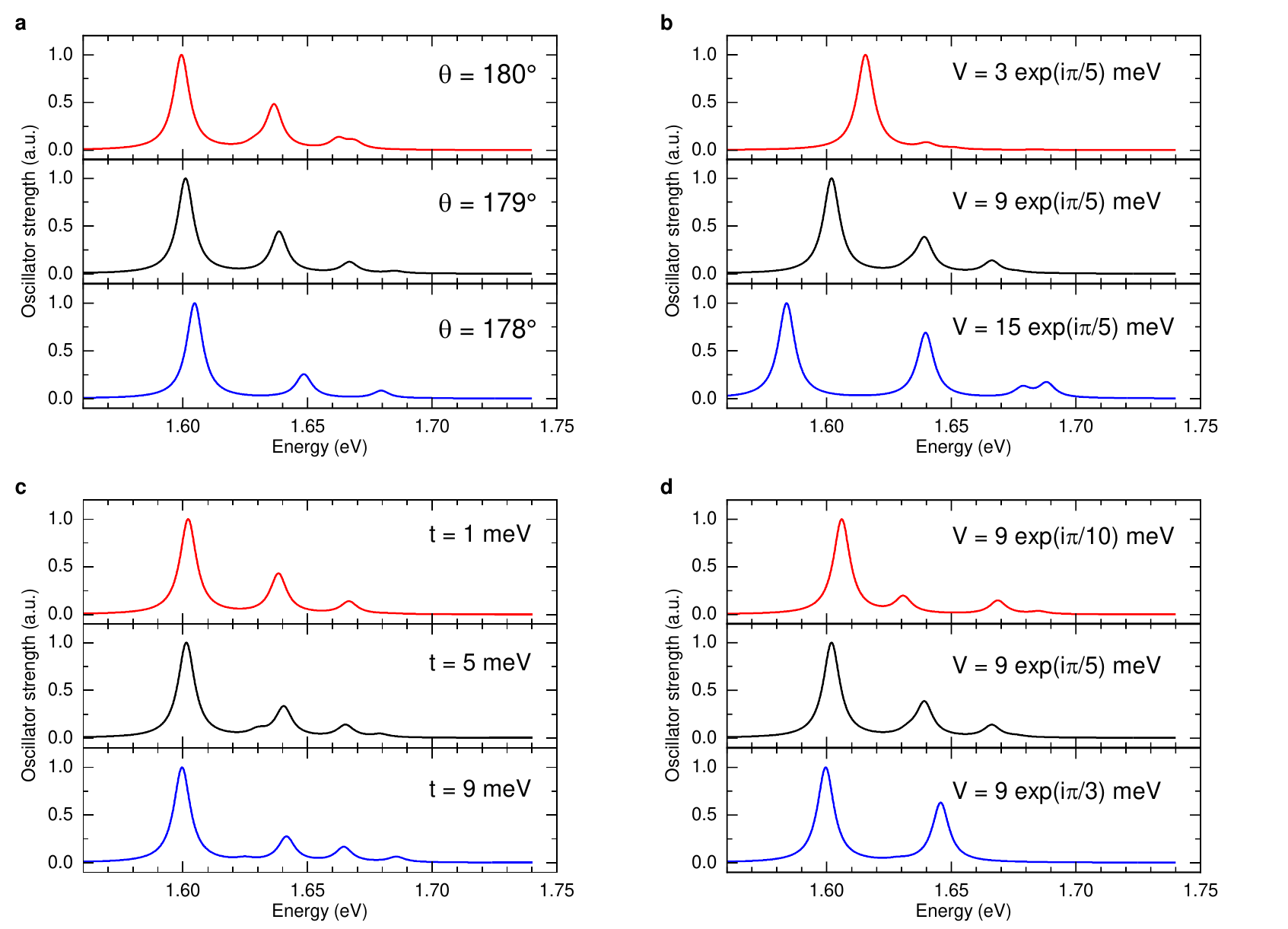}
\caption{Evolution of the computed spectra when all parameters but one are kept constant, with varying $\theta$ in (a), $|V|$ in (b), $t$ in (c) and $arg(V)$ in (d).}
\label{parameters}
\end{center}
\end{figure}

Finally, to illustrate the quantitative impact of the most important fitting parameters, we show the evolution of the computed absorption spectra with the varying parameters $\theta$, $|V|$, $arg(V)$ and $t$ in Fig.~\ref{parameters}.
%

%

\clearpage
%